\documentclass[a4paper,fleqn,10pt]{article}
\pdfoutput=1
\usepackage{amsmath}
\usepackage{amssymb}
\usepackage{array}
\usepackage{calc}
\usepackage{longtable}
\usepackage{multirow}
\usepackage{pstricks}
\usepackage{graphicx}
\usepackage{xspace}
\usepackage{units}

\numberwithin{equation}{section}
\usepackage{mciteplus}
\usepackage[a4paper,pdfborder={0 0 0}]{hyperref}
\usepackage[format=hang,labelfont=bf,hypcap=true]{caption}
\usepackage{subcaption}
\usepackage{sectsty}
\allsectionsfont{\sffamily}
\subsubsectionfont{\mdseries\itshape\large}
\makeatletter
\DeclareRobustCommand*{\bfseries}{%
  \not@math@alphabet\bfseries\mathbf
  \fontseries\bfdefault\selectfont
  \boldmath
}
\makeatother
\let\spreprint\empty
\newcommand{\preprint}[1]{\def\spreprint{\protect#1}}
\let\sinstitute\empty
\newcommand{\institute}[1]{\def\sinstitute{\protect#1}}
\makeatletter
\renewcommand{\maketitle}{\begingroup
  \null\thispagestyle{empty}%
    \ifx\spreprint\empty
      \vskip 5ex
    \else
      \flushright\large\spreprint\vskip 2ex
    \fi
    \vskip 5ex
    \flushleft
      {\sffamily\bfseries\huge\@title}\vskip 6ex
      \@author\vskip 2ex
      \ifx\sinstitute\empty
      \else
        {\small\sinstitute}
      \fi
    \vskip 5ex
  \endgroup
}
\makeatother
\renewenvironment{abstract}{\begin{center}
  {\large\sffamily\bfseries Abstract: }
  \begin{minipage}[t]{0.75\textwidth}
}{\end{minipage}\end{center}\vskip 10ex}

\numberwithin{equation}{section}
\newcommand{\MINLO}{MiN\protect\scalebox{0.8}{LO}\xspace}
\newcommand{\SMCatNLO}{S--M\protect\scalebox{0.8}{C}@N\protect\scalebox{0.8}{LO}\xspace}
\newcommand{\MCatNLO}{M\protect\scalebox{0.8}{C}@N\protect\scalebox{0.8}{LO}\xspace}

\newcommand{\POWHEG}{P\protect\scalebox{0.8}{OWHEG}\xspace}

\newcommand{\MEPSatNLO}{M\scalebox{0.8}{E}P\scalebox{0.8}{S}@N\scalebox{0.8}{LO}\xspace}
\newcommand{\CKKW}{CKKW\xspace}
\newcommand{\UNLOPS}{UN\scalebox{0.8}{LO}P\scalebox{0.8}{S}\xspace}

\newcommand{\Rivet}{R\protect\scalebox{0.8}{IVET}\xspace}

\newcommand{\MCFM}{M\protect\scalebox{0.8}{CFM}\xspace}
\newcommand{\Sherpa}{S\protect\scalebox{0.8}{HERPA}\xspace}

\newcommand{\Amegic}{A\protect\scalebox{0.8}{MEGIC++}\xspace}

\long\def\symbolfootnote[#1]#2{\begingroup%
\def\thefootnote{\fnsymbol{footnote}}\footnote[#1]{#2}\endgroup}

\newcommand{\abs}[1]{\left| #1\right|}

\newcommand{\abr}[1]{\langle #1\rangle}

\newcommand{\sbr}[1]{\left[ #1\right]}
\newcommand{\im}{\imath}
\newcommand{\jm}{\jmath}

\newcommand{\done}{{\rm d}}
\newcommand{\order}{\mathcal{O}}
\newcommand{\mc}[1]{\mathcal{#1}}
\newcommand{\mr}[1]{\mathrm{#1}}

\newcommand{\dst}{\displaystyle}

\newcommand{\bea}{\begin{eqnarray}}
\newcommand{\eea}{\end{eqnarray}}
\newcommand{\bi}{\begin{itemize}}
\newcommand{\ei}{\end{itemize}}

\newcommand{\Hl}{\vphantom{$\int\limits_a^b$}}
\newcommand{\reserrs}[3]{$#1_{-#2}^{+#3}$ pb}

\hypersetup{
  pdfauthor={Stefan Hoeche, Frank Krauss, Marek Schoenherr},
  pdftitle={Uncertainties in MEPS@NLO calculations of h+jets}
}
\preprint{SLAC-PUB 15899\\IPPP/14/08\\DCPT/14/16\\LPN14-009\\MCNET-14-02}
\author{Stefan H{\"o}che$^1$, Frank Krauss$^2$, Marek Sch{\"o}nherr$^2$}
\title{\scalebox{0.95}{Uncertainties in \MEPSatNLO calculations of $h+$jets}}
\institute{$^1$ SLAC National Accelerator Laboratory, 
  Menlo Park, CA 94025, USA\\
  $^2$ Institute for Particle Physics Phenomenology,
  Durham University, Durham DH1 3LE, UK}
\begin{document}
\maketitle
\begin{abstract}
  Uncertainties in the simulation of Higgs boson production with
  up to two jets at next-to leading order accuracy are investigated.  
  Traditional uncertainty estimates based on scale variations are extended 
  employing different functional forms for the central scale, and 
  the impact of details in the implementation of the parton shower
  is discussed.  
\end{abstract}
\section{Introduction}
\label{sec:intro}
After the discovery of a Standard-Model like Higgs boson at Run I of the LHC 
operations~\cite{Aad:2012tfa,*Chatrchyan:2012ufa} further studies of the 
properties of the new particle will become a focal point of the 
physics programme during Run II.  Anticipating the effect of the scheduled 
significantly larger collision energy of the protons translating into 
increased energies available for colliding partons, and a vastly increased 
luminosity, more production and decay channels and combinations of both become
available for studying the coupling of the Higgs boson to
other particles.  In particular, production channels such as the production
of the Higgs boson in weak boson fusion, yielding two additional jets, or its
production in the boosted regime will provide challenging tests for
the Brout-\-Englert-\-Higgs mechanism of mass generation in a gauge invariant
way~\cite{Englert:1964et,*Higgs:1964pj,*Guralnik:1964eu,*Kibble:1967sv}.  

With increasing accuracy in experimental measurements, theoretical computations
are inevitably required to become more precise.  For the dominant production 
of the Higgs boson through gluon-induced heavy quark loops, tremendous 
progress has been made over the past years to achieve the necessary precision, 
for example with calculations at the next-to-\-next-to-\-leading order accuracy 
(NNLO) in the perturbative expansion of the strong coupling: In the past year
the by now routinely employed result for $pp\to h$ through gluon fusion in the 
effective theory~\cite{Melnikov:2006di,*Grazzini:2008tf} has been supplemented 
with a calculation for $pp\to h+{}$jet~\cite{Boughezal:2013uia}.  Mass effects in
the heavy quark loop, going beyond the usual effective theory approach, were 
evaluated in~\cite{Harlander:2012hf}.  Even a complete N$^3$LO calculation of 
$pp\to h$ is being finalized at the moment, with first results already being 
reported~\cite{Anastasiou:2013mca}.  Mixed QCD and electroweak two-\-loop 
corrections have been evaluated in two different 
ways~\cite{Anastasiou:2008tj,*Actis:2008ug}.  At NLO accuracy, due to the level
of automation achieved by now, the production of the Higgs boson in association
with two~\cite{Campbell:2006xx,*Campbell:2010cz,*vanDeurzen:2013rv} and three 
jets~\cite{Cullen:2013saa} through the effective vertex has been investigated.
Due to the gluon initial states, resummation plays an important role.  
Results for the transverse momentum in 
inclusive production at NNLO+NNLL accuracy, are available for example 
in~\cite{deFlorian:2009hc,*deFlorian:2012mx}.  Jet vetoes, which are 
particularly relevant for the Higgs boson decaying into $W$ bosons have been 
discussed in~\cite{Banfi:2012yh,*Banfi:2012jm,
  *Tackmann:2012bt,*Stewart:2013faa,*Becher:2012qa,*Becher:2012yn,
  *Becher:2013xia,*Liu:2013hba}.  

Parton-\-shower based simulations, typically used by the experiments, usually 
lag behind analytical calculations by at least one perturbative order.  This 
is exemplified by the highest accuracy in simulating this process so far, 
which achieves the NNLO level through a suitable reweighting in 
the \MINLO procedure~\cite{Hamilton:2013fea}.  
At the next-to leading order accuracy, parton-\-shower matched calculations 
have been provided using the \POWHEG~\cite{Nason:2004rx,*Frixione:2007vw,
  *Alioli:2008tz,*Hamilton:2009za,*Hoche:2010pf,*Campbell:2012am} and 
\MCatNLO~\cite{Frixione:2002ik,*Frixione:2010ra} methods.  Recently, the 
\MCatNLO algorithm has been modified in order to provide the full color-correct
NLO result~\cite{Hoeche:2011fd,*Hoeche:2012ft,*Hoeche:2012fm} (\SMCatNLO), 
which also forms the basis of a multijet merging algorithm based on NLO 
calculations~\cite{Hoeche:2012yf,*Gehrmann:2012yg}.  This merging method 
is similar in spirit to the by now traditional multijet merging for LO matrix 
elements~\cite{Catani:2001cc,*Krauss:2002up,Lonnblad:2001iq,Hoeche:2009rj}.
The new NLO algorithm was successfully applied to a number of relevant physics 
cases, most notably to jet vetoes in $W$--pair backgrounds to Higgs boson 
production at the LHC~\cite{Cascioli:2013gfa} and to top-quark pair production
and the asymmetries encountered at the Tevatron~\cite{Hoeche:2013mua}.  A 
closely related multijet merging algorithm has also been proposed 
in~\cite{Lonnblad:2012ix}, while~\cite{Frederix:2012ps} relies on ideas
in line with the MLM multijet merging prescription for leading order 
calculations~\cite{Mangano:2006rw,*Alwall:2007fs}. 

For the sub-\-dominant production of a Higgs boson through weak vector boson 
fusion (VBF)~\cite{Rainwater:1997dg,*Rainwater:1998kj,*Plehn:1999xi}, which 
becomes a very interesting channel when jet vetoes between the two relatively
forward tagging jets are applied~\cite{Dokshitzer:1987nc,*Dokshitzer:1991he}, 
NLO QCD corrections in the structure function approach assume a relatively 
simple form.  They have been known for a long 
time~\cite{Han:1992hr,*Spira:1997dg,*Figy:2003nv,*Figy:2004pt,*Berger:2004pca}.
NNLO corrections in this approach have more recently been computed 
in~\cite{Bolzoni:2010xr,*Bolzoni:2011cu}, while QCD NLO corrections for $pp\to h+3$ jet
production in VBF were first discussed in~\cite{Figy:2007kv}.  The electroweak 
NLO corrections to VBF Higgs boson production~\cite{Ciccolini:2007ec} were 
found to be of roughly the same size as the QCD ones.  On the level of 
parton--\-shower based simulations, this topology is available through both 
the \POWHEG and the \MCatNLO 
algorithms~\cite{Nason:2009ai,*D'Errico:2011sd,*Frixione:2013mta}.

In this publication, a next-to-leading order plus parton shower merged 
calculation is presented for the production of a Higgs boson through gluon 
fusion in the effective theory approximation with up to two additional jets 
at NLO and a third jet at LO accuracy.  For these studies the Monte Carlo 
event generator \Sherpa~\cite{Gleisberg:2003xi,*Gleisberg:2008ta} is used
in conjunction with virtual corrections taken from \MCFM~\cite{Campbell:2006xx,
  *Campbell:2010cz,Campbell:2010ff,*MCFM}.  The focus of the study
rests on the accurate description both of the rate and of those kinematical 
distributions of the various objects in the final state, which are central to 
analyses involving typical VBF cuts.  The fully inclusive nature of a 
Monte Carlo simulation using a general-purpose event generator ensures, however,
that a wide range of inclusive and exclusive observables can be analyzed
simultaneously.  The most relevant uncertainties in the simulation are 
detailed and some substantial differences between one-jet and two-jet 
NLO-merged simulations are pointed out.  Special attention is paid to
uncertainties related to the functional form of scale in the fixed-order 
NLO calculation, which massively exceed those obtained from a mere variation 
of a constant scale factor in the conventional range from $1/2$ to $2$.  

This manuscript is organized as follows: Section~\ref{sec:methods} reviews 
the methods used in the simulation, with emphasis on the actual multijet
merging algorithm.  Section~\ref{sec:results} presents predictions obtained 
with the event generator \Sherpa and discusses related uncertainties.  
An outlook is given in Sec.~\ref{sec:conclusion}.

\section{Methods}
\label{sec:methods}

This section briefly summarizes the \SMCatNLO matching method and the \MEPSatNLO
merging technique. We emphasize only those aspects of the implementation in
\Sherpa which are relevant to the assessment of uncertainties of the simulation.
Details of the two algorithms are described 
in~\cite{Hoeche:2011fd,*Hoeche:2012ft,*Hoeche:2012fm}
and~\cite{Hoeche:2012yf,*Gehrmann:2012yg}.

\subsection{Matrix-element parton-shower matching}
\label{sec:mcatnlo}

The action of the parton shower on an arbitrary parton-level final state can be
expressed in terms of a generating functional, $\mc{F}_n(t)$, where $n$ is the
number of existing partons, and $t$ is the parton shower starting scale.
The value of an infrared safe observable, $O$, in the Born approximation is
then computed as
\begin{equation}\label{eq:match_lo}
  \abr{O}^{\rm(PS)}=\int\done\Phi_B\,\mr{B}(\Phi_B)\,\mc{F}_{0}(\mu_F^2,O)\;,
\end{equation}
with $\done\Phi_B$ the differential Born phase space element and 
$\mr{B}(\Phi_B)$ the Born differential cross section. The generating 
functional of the parton shower reads
\begin{equation}\label{eq:gen_ps}
  \mc{F}_n(t,O)\,=\;
  \Delta_n(t_c,t)\,O(\Phi_n)
  +\int_{t_c}^t\done\Phi_1^\prime\,
                      \mr{K}_n(\Phi_1^\prime)\,\Delta_n(t^\prime,t)\,
                      \mc{F}_{n+1}(t^\prime,O)
\end{equation}
with $\mr{K}_n(\Phi_1)$ the parton shower splitting kernel on the $n$ 
parton state and $\Delta_n(t^\prime,t)$ is the corresponding Sudakov form factor. 
The single parton emission phase space is parametrized as $\done\Phi_1=\done t\,
\done z\,\done\phi\,J(t,z,\phi)$. In this context $t\equiv t(\Phi_1)$ is the 
evolution variable, $z$ is the splitting variable, $\phi$ is the 
azimuthal angle of the splitting and $J(t,z,\phi)$ is the associated 
Jacobian. $t_c$ is the infrared cutoff.
While the first term in Eq.~\eqref{eq:gen_ps} describes the 
no-emission probability, the second term describes a single independent 
emission at scale $t^\prime$ including the ensuing iteration with the 
boundary conditions of the newly formed state.

The \MCatNLO matching method promotes Eq.~\eqref{eq:match_lo} to NLO accuracy
using a modified subtraction scheme~\cite{Frixione:2002ik}. This technique 
was extended in \cite{Hoeche:2011fd,*Hoeche:2012ft,*Hoeche:2012fm} such that the first
emission in the shower is generated in a fully coherent manner, and therefore
all singularities of the real-emission matrix element are properly subtracted.
This applies in particular to terms which are suppressed by $1/N_c$.
We will refer to the latter method as \SMCatNLO.

In \SMCatNLO, any observable $O$ is computed as
\begin{equation}\label{eq:mcatnlo}
  \abr{O}^\text{(\SMCatNLO)}\,=\;\int\done\Phi_B\,\bar{\mr{B}}^{\rm(A)}(\Phi_B)\,
    \mc{F}^{\rm(A)}(\mu_Q^2,O)
  +\int\done\Phi_R\,\mr{H}^{\rm(A)}(\Phi_R)\,\mc{F}_1(t,O)\;,
\end{equation}
where $\bar{\mr{B}}^{\rm(A)}$ and $\mr{H}^{\rm(A)}$ are the next-to-leading 
order weighted differential cross section and the hard remainder 
function, given by
\begin{equation}\label{eq:mcatnlo_bbar_h}
  \begin{split}
  \bar{\mr{B}}^{\rm(A)}(\Phi_B)\,=&\;\mr{B}(\Phi_B)+\tilde{\mr{V}}(\Phi_B)
    +\mr{I}^{\rm(S)}(\Phi_B)+\int\done\Phi_1\,\Big[\,
      \mr{D}^{\rm(A)}(\Phi_B,\Phi_1)\,\Theta\left(\mu_Q^2-t\right)
      -\mr{D}^{\rm(S)}(\Phi_B,\Phi_1)\,\Big]\;,\\
  \mr{H}^{\rm(A)}(\Phi_R)\,=&\;\mr{R}(\Phi_R)
  -\mr{D}^{\rm(A)}(\Phi_R)\,\Theta\left(\mu_Q^2-t\right)\;.
  \end{split}
\end{equation}
Here we have introduced the virtual corrections, $\tilde{\mr{V}}(\Phi_B)$,
the real-emission corrections, $\mr{R}(\Phi_R)$, and the corresponding
real-emission phase space element $\done\Phi_R$. Most importantly, 
the dipole subtraction terms are given by $\mr{D}(\Phi_B,\Phi_1)$ in 
unintegrated form, and by $\mr{I}(\Phi_B)$ in integrated form. 
We distinguish between fixed-order subtraction terms, $\mr{D}^{\rm(S)}$ 
and shower subtraction terms, $\mr{D}^{\rm(A)}$. Both must have the same 
kinematics mapping, however their functional form away from the singular limits 
of the real emission corrections may differ. In particular, we implement an 
upper cutoff in the parton-shower evolution parameter in $\mr{D}^{\rm(A)}$ only, 
which is referred to as the resummation scale $\mu_Q$.

The differential real-emission phase space element factorizes as 
$\done\Phi_R = \done\Phi_B\,\done\Phi_1$ with the above parametrization
for $\done\Phi_1$. This factorization allows to define a generating functional 
$\mc{F}^{\rm(A)}(t,O)$ of the \SMCatNLO as
\begin{equation}\label{eq:gen_mcatnlo}
  \mc{F}^{\rm(A)}(t,O)\,=\;
    \Delta^{\rm(A)}(t_c,t)\,O(\Phi_B)
    +\int_{t_c}^{t}\!\done\Phi_1^\prime\,
      \frac{\mr{D}^{\rm(A)}(\Phi_B,\Phi_1^\prime)}{\mr{B}(\Phi_B)}\,
      \Delta^{\rm(A)}(t^\prime,t)\,\mc{F}_1(t^\prime,O)
\end{equation}
with $\Delta^{\rm(A)}(t_c,t)$ the Sudakov factor of the \SMCatNLO,
and $\mr{D}^{\rm(A)}(\Phi_B,\Phi_1^\prime)/\mr{B}(\Phi_B)$ its splitting
kernels. Again, the first term in Eq.~\eqref{eq:gen_mcatnlo} is simply a 
no-emission probability, while the second term now describes one fully
coherent emission from the \SMCatNLO. Any further radiation is
implemented in the parton-shower approximation, as indicated by 
$\mc{F}_1(t',O)$.

We employ a parton shower based on Catani-Seymour dipole 
subtraction~\cite{Schumann:2007mg}. To assess the uncertainty 
arising from the choice of evolution variable, we implement two
different options, which are listed in Tab.~\ref{tab:evol_vars}.
Their impact on experimental observables is analyzed in 
Sec.~\ref{sec:results}.

\begin{table}
  \begin{center}
    \renewcommand{\arraystretch}{1.25}
    \begin{tabular}{lcc}
      \hline
      Scheme & Final State & Initial State\\\hline
      0 & 
      $\dst 2\,p_ip_j\,\tilde{z}_{i,jk}(1-\tilde{z}_{i,jk})$ &
      $\dst 2\,p_ap_j\,(1-x_{aj,k})$ \\[2mm]
      1 &
      $\dst 2\,p_ip_j\,\left\{\begin{array}{cl}
        \tilde{z}_{i,jk}(1-\tilde{z}_{i,jk}) & \text{if $i,j=g$}\\
        1-\tilde{z}_{i,jk} & \text{if $j=g$}\\
        \tilde{z}_{i,jk} & \text{if $i=g$}\\
        1 & \text{else}\\
      \end{array}\right.$ &
      $\dst 2\,p_ap_j\,\left\{\begin{array}{cl}
        1-x_{aj,k} & \text{if $j=g$}\\
        1 & \text{else}\\
      \end{array}\right.$ \\[10mm]
      \hline
    \end{tabular}
    \caption{Evolution parameters for the parton shower.
      We use the variables defined in~\cite{Catani:1996vz}.
      \label{tab:evol_vars}}
  \end{center}
\end{table}

In addition, two different kinematics mappings are implemented, which were 
described and compared in detail in~\cite{Hoeche:2009xc,*Carli:2009cg}. 
We denote the original mapping, proposed in~\cite{Schumann:2007mg} and derived 
from~\cite{Catani:1996vz} by ``scheme 1'', while the mapping which generates 
recoil also in initial state splittings with final state spectators is denoted 
as ``scheme 0''.  This scheme can be described as 
follows~\cite{Hoeche:2009xc,*Carli:2009cg}:

Particle momenta after the splitting process $\widetilde{\im\jm}\to ij$ 
in the presence of a spectator $k$ are expressed in terms of the original 
momenta, $\tilde{p}_{ij}$ and $\tilde{p}_k$, as~\footnote{
  We work in the five--\-flavor scheme and therefore consider massless partons 
  only.  All momenta are taken as outgoing.}
\begin{align}\label{eq:def_pi_pj_FF}
  p_i^\mu\,=&\;z_{i,jk}\,\tilde{p}_{ij}^\mu+\frac{{\rm k}_\perp^2}{z_{i,jk}}\,
    \frac{\tilde{p}_k^\mu}{2\,\tilde{p}_{ij}\tilde{p}_k}+k_\perp^\mu\;,
  &p_j^\mu\,=&\;(1-z_{i,jk})\,\tilde{p}_{ij}^\mu+\frac{{\rm k}_\perp^2}{1-z_{i,jk}}\,
    \frac{\tilde{p}_k^\mu}{2\,\tilde{p}_{ij}\tilde{p}_k}-k_\perp^\mu\;,
\end{align}
where
\begin{equation}\label{eq:def_zi_kt}
  \begin{split}
    {\rm k}_\perp^2\,=&\;2\,\tilde{p}_{ij}\tilde{p}_k\;y_{ij,k}\;z_{i,jk}\,(1-z_{i,jk})\;.
  \end{split}
\end{equation}
The parameters $z_{i,jk}$ and $y_{ij,k}$ depend on the type of splitting and
are given for all dipole configurations in Tab.~\ref{tab:def_eta_xi}. 
The spectator momentum $p_k$ is determined by momentum conservation.
In the case of initial state splitter or spectator partons, a proper Lorentz
transformation is applied to keep both initial state particles aligned along
the beam axis.

In scheme 1, initial-state splittings with final state spectator are instead 
constructed as if they were final-state splittings with initial state spectator,
by replacing $\tilde{z}_j\to u_j$, $p_k\to -p_i$ and $p_j\to p_k$. 

\begin{table}
  \begin{center}
    \renewcommand{\arraystretch}{1.25}
    \begin{tabular}{ccccccc}
      \cline{1-3}\cline{5-7}
      Configuration &
      \hspace*{9mm}$z_{i,jk}$\hspace*{9mm} & 
      \hspace*{7mm}$y_{ij,k}$\hspace*{7mm} & &
      Configuration & 
      \hspace*{9mm}$z_{j,ak}$\hspace*{9mm} & 
      \hspace*{7mm}$y_{ja,k}$\hspace*{7mm}\\
      \cline{1-3}\cline{5-7}\\[-3mm]
      Final-Final & $\tilde{z}_i$ & $\dst y_{ij,k}$ & &
      Initial-Final & $\dst\frac{x_{jk,a}-1}{x_{jk,a}-u_j}$ &
      $\dst\frac{u_j}{x_{jk,a}}$ \\[3mm]
      Final-Initial & $-\,\tilde{z}_i$ & $x_{ij,a}-1$ & &
      Initial-Initial & $\dst1-\frac{1}{x_{j,ab}+\tilde{v}_j}$ &
      $\dst\frac{-\,\tilde{v}_j}{x_{j,ab}}$ \\[3mm]
      \cline{1-3}\cline{5-7}
    \end{tabular}
    \caption{Mapping of variables for parton shower kinematics in
      scheme 0, see Eqs.~\eqref{eq:def_pi_pj_FF} and~\eqref{eq:def_zi_kt}.
      \label{tab:def_eta_xi}}
  \end{center}
\end{table}

\subsection{Multijet merging}
\label{sec:mepsatnlo}

The \SMCatNLO method augments NLO fixed-order calculations with 
the simple resummation encoded in the parton shower. Compared to
pure fixed-order calculations, this allows more meaningful 
definitions of jet cross sections, because logarithms of the 
jet transverse momentum cutoff are resummed to (at least) 
leading logarithmic accuracy. The precision of such a simulation 
can be extended further by correcting emissions from the parton shower 
with fixed-order results for higher jet multiplicity. This has been
achieved both at leading order~\cite{Hamilton:2010wh,Hoeche:2010kg} 
and at next-to-leading order~\cite{Hoeche:2012yf,*Gehrmann:2012yg,
  Lonnblad:2012ix,Frederix:2012ps}. While the merging at LO is 
straightforward, it requires additional care at NLO to properly
subtract the first-order expansion of the shower expression.
This procedure is vital to preserve the logarithmic accuracy
of the parton shower~\cite{Hoeche:2012yf,*Gehrmann:2012yg,Lonnblad:2012ix}.

It is instructive to analyze the contribution to an observable $O$
from the exclusive simulation of final states with $n$ hard partons.
Additional partons may be present, which are not resolved 
according to a technical jet cut, $Q_{\rm cut}$, called 
the merging cut. The observable $Q$ in which the merging cut is specified
is called the merging criterion. It may or may not be identical to an
experimental jet definition, however, to make the calculation useful 
in practice, one should choose the two as similar as possible.

The exclusive contribution to the observable with exactly 
$n$ hard partons reads
\begin{equation}\label{eq:nlo_term}
  \begin{split}
    \abr{O}_{n}^{\rm excl}
    \,=&\;\int\done\Phi_{n}\,\Theta(Q(\Phi_{n})-Q_{\rm cut})\;
          \tilde{\mr{B}}_{n}^\text{(A)}(\Phi_n)\,
          \tilde{\mc{F}}^{\rm(A)}_{n}(\mu_Q^2,O\,;<\!Q_{\rm cut})\\
    &+\int\done\Phi_{n+1}\,\Theta(Q(\Phi_{n})-Q_{\rm cut})\,
         \Theta(Q_{\rm cut}-Q(\Phi_{n+1}))\,
         \tilde{\mr{H}}_{n}^\text{(A)}(\Phi_{n+1})\,
         \tilde{\mc{F}}_{n+1}(\mu_Q^2,O\,;<\!Q_{\rm cut})\;,
  \end{split}
\end{equation}
In this context, we have introduced a new generating functional, 
$\tilde{\mc{F}}_{n}(\mu_Q^2,O\,;<\!Q_{\rm cut})$, which represents
a truncated vetoed parton shower~\cite{Nason:2004rx,*Frixione:2007vw,Hoeche:2009rj}.
It implements emissions on a parton shower tree that corresponds 
to the $n$-parton final state. It also computes the survival probability
for that particular configuration, indicated by the notation $<\!Q_{\rm cut}$. 
The starting conditions must be chosen carefully in order to not to spoil 
the accuracy of the calculation: Each possible shower topology is selected 
according to the exact forward branching probability of the shower into 
the given configuration. This scheme was described in great detail 
in~\cite{Lonnblad:2001iq,Hoeche:2009rj,Lonnblad:2012ix}.

The seed cross sections $\bar{\mr{B}}^{\rm(A)}$ and $\mr{H}^{\rm(A)}$
as defined in Eq.~\eqref{eq:mcatnlo_bbar_h} are replaced by the functions 
\begin{equation}\label{eq:mcatnlon_bbar_h}
  \begin{split}
    \tilde{\rm B}_{n}^{\rm (A)}(\Phi_{n})=&\;
      {\rm B}_{n}(\Phi_{n})+\tilde{\mr{V}}_{n}(\Phi_{n})
        +{\rm I}_{n}^{\rm (S)}(\Phi_{n})
      +\int\done\Phi_1\,\sbr{\tilde{\rm D}_{n}^{\rm (A)}(\Phi_n,\Phi_1)
        -{\rm D}_{n}^{\rm (S)}(\Phi_n,\Phi_1)}\\
    \tilde{\mr{H}}_{n}^{\rm(A)}(\Phi_{n+1})=&\;
    \mr{R}_{n}(\Phi_{n+1})-
      \tilde{\mr{D}}_{n}^{\rm(A)}(\Phi_{n+1})\;,\\
  \end{split}
\end{equation}
which take the probability of truncated parton shower emissions into
account~\cite{Hoeche:2012yf,*Gehrmann:2012yg}. To this end, the dipole
terms used in the \SMCatNLO are extended by the parton-shower emission
probabilities, $\mr{B}_{n}(\Phi_n)\,\mr{K}_{i}(\Phi_{1,n+1})$, where
$\mr{K}_{i}(\Phi_{1,n+1})$ is the sum of all shower splitting functions
for the intermediate state with $i<n$ in the predefined shower tree.
\begin{equation}\label{eq:compound_kernel}
   \tilde{\mr{D}}_{n}^{\rm(A)}(\Phi_{n+1})\,=\;
   \mr{D}_{n}^{\rm(A)}(\Phi_{n+1})\,\Theta(t_{n}-t_{n+1})\;
   +\sum_{i=0}^{n-1}\mr{B}_{n}(\Phi_n)\,\mr{K}_{i}(\Phi_{1,n+1})\,
      \Theta(t_i-t_{n+1})\,\Theta(t_{n+1}-t_{i+1})\,\Big|_{\,t_0=\mu_Q^2}\;.
\end{equation}
While seemingly quite complex, Eq.~\eqref{eq:compound_kernel} has a 
very simple physical interpretation: The first term corresponds to
the coherent emission of a parton from the external $n$-parton final 
state. It contains all soft and collinear singularities which are
present in the real-emission matrix elements. The sum in the second term
corresponds to emissions from the intermediate states with $i$ partons.
They can be implemented in the parton shower approximation, because 
soft divergences are regulated by the finite mass of the intermediate
particles.

In practice the second term in Eq.~\eqref{eq:compound_kernel} can be
implemented in an NLO-vetoed truncated shower~\cite{Hoeche:2012yf,*Gehrmann:2012yg}.
This is a straightforward modification of an existing shower algorithm.
It is far more complicated to select the correct renormalization scale.
Two different methods have been proposed to this end, which are both constructed
such that the logarithmic accuracy of the shower is maintained. Their results
therefore differ only beyond NLO.
\begin{itemize}
\item The \UNLOPS scale~\cite{Lonnblad:2012ix}\\
An arbitrary scale is chosen for the fixed-order NLO calculation. 
Renormalization terms and collinear mass factorization counterterms
are added to restore the scale choice of the parton shower at NLO accuracy. 
The procedure recovers all logarithms formally resummed by the parton shower,
but it does not reproduce its full logarithmic structure as the two-loop
running of the strong coupling generates additional terms.
\item The \CKKW scale~\cite{Hoeche:2012yf,*Gehrmann:2012yg}\\
The scale in the fixed-order calculation is chosen such that the 
coupling factors of the parton shower for the chosen shower history
are recovered entirely. Renormalization terms are then exactly zero 
and only collinear mass factorization counterterms must be added.
\end{itemize}
We will compare the effects of these two choices in Sec.~\ref{sec:results},
with some different parametrization used for the \UNLOPS scale.

\section{Results}
\label{sec:results}

This section presents results obtained with the \MEPSatNLO algorithm
applied to Higgs boson production through gluon 
fusion in association with jets at a center-of-mass energy of 8 TeV. 
We work in the five flavor scheme. 
The loop-mediated coupling of the Higgs boson to gluons is calculated 
in Higgs effective theory (HEFT)~\cite{Ellis:1975ap,*Wilczek:1977zn,
  *Shifman:1979eb,*Ellis:1979jy} with $m_t\to\infty$. 
The Born- and real-emission matrix elements as well as the dipole 
subtraction terms~\cite{Catani:1996vz} are computed using 
\Amegic~\cite{Krauss:2001iv,*Gleisberg:2007md}.
One-loop matrix elements are implemented according 
to~\cite{Dawson:1990zj,*Djouadi:1991tka} in the case of $pp\to h$ 
and~\cite{deFlorian:1999zd,*Ravindran:2002dc} in the case of $pp\to h+{}$jet, 
or obtained through an interface to 
\MCFM~\cite{Campbell:2006xx,*Campbell:2010cz,Campbell:2010ff,*MCFM}
in the case of $pp\to h+2$ jets.
Our calculation is purely perturbative, i.e.\ hadronization and 
underlying event contributions are not included. 
The CT10nlo~\cite{Lai:2010vv,*Gao:2013xoa} parton distribution 
functions are used.

The Higgs boson mass is set to $m_h=125$~GeV. As we are interested in the 
properties of the QCD activity accompanying the production of the Higgs boson,
no restrictions on its decay are applied. Jets are defined using the anti-$k_\perp$ 
jet algorithm~\cite{Cacciari:2008gp} with $R=0.4$ and $p_\perp^\text{min}=30$ GeV.
Jets are required to have a rapidity of $\abs{y}<5$.
A VBF selection is defined by requiring at least two jets and imposing 
two additional cuts on the two leading-$p_T$ ones:
$\abs{\Delta y_{j_1,j_2}}>2.8$ and $m_{j_1,j_2}>400$~GeV, and, where 
indicated, veto the emission of a third jet inbetween the two leading 
jets.

Uncertainties of the perturbative calculation
arise from a variation of all unphysical scales in the process. 
While the factorization and renormalization scales, $\mu_F$ 
and $\mu_R$ are varied independently within the conventional factor of two, the 
resummation scale $\mu_Q$ is varied by a factor of $\sqrt{2}$. The 
merging scale $Q_\text{cut}$, separating $pp\to h+n$ jet from  
$pp\to h+(n+1)$ jet calculations, is varied in the set $\{15,20,30\}$ 
GeV\footnote
{
  The merging scale should not be set to a value above the minimum 
  transverse momentum of the analysis jet definition in order not to 
  degrade the accuracy of the perturbative description of observables 
  involving at least one jet. This avoids muddying the merging scale 
  systematics with leading order vs.\ next-to-leading order effects.
}. 
The central scale choices are $\mu_{R}=\mu_\text{CKKW}=\alpha_s^2(m_h)\,\alpha_s(t_1)
\cdots\alpha_s(t_n)$, and $\mu_F=\mu_Q=m_h$. To assess the large relative 
$\order(\alpha_s^2)$ effects in this process, which arise from the
Higgs effective coupling to gluons, two additional functional forms 
of the renormalization scale are investigated using the \UNLOPS method: 
$\mu_R=m_h$ and $\mu_R=\hat{H}_T^\prime=\sum m_\perp$ 
(sum of all transverse masses in the final state)~\cite{Berger:2010zx}. 
The logarithmic accuracy of the parton shower is preserved 
in this case by including the renormalization term~\cite{Hoeche:2012yf}
\begin{equation}
  {\rm B}_{n}\,\frac{\alpha_s(\mu_R)}{\pi}\,\beta_0\,
  \left(\log\frac{\mu_R}{\mu_\text{CKKW}}\right)^{2+n}
\end{equation}
This term reverts any scale choice to that of $\mu_\text{CKKW}$ 
to one-loop order, cf.\ Sec.~\ref{sec:methods}.
We will choose the prediction for $\mu_R=m_h$ as a reference for comparisons.
The \CKKW scale and $\hat{H}_T^\prime$ can be regarded as two extreme
choices on opposite sides of $m_h$. While $\hat{H}_T^\prime$ increases 
in the presence of an additional low-$p_T$ jet, the \CKKW scale decreases.

All analyses and plots in this section were made with the help of
\Rivet~\cite{Buckley:2010ar}.

\subsection{Inclusive observables}
\label{sec:res-incl}

\begin{table}[t]
  \begin{center}
    \begin{tabular}{|l|c|c|c|}
      \hline
      \Hl & $\mu_R=\mu_\text{CKKW}$ & $\mu_R=m_h$ & $\mu_R={\hat H}_{\rm T}^\prime$ \\
      \hline
      $\sigma_\text{0 jet}^\text{incl}$\Hl & \reserrs{12.2}{1.5}{1.6} & \reserrs{11.6}{1.4}{1.7} & \reserrs{10.9}{1.2}{1.4} \\
      \hline
      $\sigma_\text{0 jet}^\text{excl}$\Hl & \reserrs{8.05}{0.82}{0.82} & \reserrs{7.71}{0.85}{0.96} & \reserrs{7.37}{0.76}{0.86} \\
      \hline
      $\sigma_\text{1 jet}^\text{incl}$\Hl & \reserrs{4.16}{0.73}{0.91} & \reserrs{3.91}{0.54}{0.78} & \reserrs{3.54}{0.48}{0.63} \\
      \hline
      $\sigma_\text{1 jet}^\text{excl}$\Hl & \reserrs{3.08}{0.51}{0.48} & \reserrs{2.92}{0.42}{0.50} & \reserrs{2.68}{0.39}{0.46} \\
      \hline
      $\sigma_\text{2 jet}^\text{incl}$\Hl & \reserrs{1.07}{0.22}{0.46} & \reserrs{0.99}{0.13}{0.30} & \reserrs{0.86}{0.10}{0.18} \\
      \hline
      $\sigma_\text{VBF cuts}$\Hl & \reserrs{0.165}{0.039}{0.070} & \reserrs{0.152}{0.021}{0.043} & \reserrs{0.126}{0.014}{0.021} \\
      \hline
      $\sigma_\text{VBF cuts}^\text{central jet veto}$\Hl & \reserrs{0.124}{0.028}{0.048} & \reserrs{0.113}{0.017}{0.031} & \reserrs{0.096}{0.013}{0.018} \\
      \hline
    \end{tabular}
  \end{center}
  \caption{
           Cross sections with three different central scales.
           Uncertainties are given as super- and subscripts and include
           $\mu_{R/F}$ variations, $\mu_Q$ variations, and $Q_{\rm cut}$ variations.
           \label{tab:xsecs}}
\end{table}

First we examine inclusive and exclusive (jet) cross sections 
generated in our simulations. They are summarized in Tab.~\ref{tab:xsecs}. 
Although the expressions for the respective quantities are all calculated 
at next-to-leading order accuracy and only differ 
by terms of $\order(\alpha_s^2)$, induced through the different scales 
used for the strong coupling, the deviations from one another
are rather large already. This hints at the well known fact of 
the poor convergence of the perturbative series in this process. Generally, 
the computed cross sections are largest with $\mu_R=\mu_\text{CKKW}$ and 
smallest with $\mu_R={\hat H}_{\rm T}^\prime$, increasingly so as the 
hardness of the event increases as required by the respective selection 
criteria. The inclusive cross section, which is expected to be less 
sensitive to such kinematic effects, is in good agreement between the 
different scale choices.

The uncertainties quoted stem from a variation of $\mu_F$ and $\mu_R$, 
$\mu_Q$, and $Q_\text{cut}$, varied separately as detailed above and then 
summed in quadrature. Again, the different choices for the central scale 
lead to increasingly deviating uncertainty estimates, smallest for 
$\mu_R={\hat H}_{\rm T}^\prime$ and largest for $\mu_R=\mu_\text{CKKW}$.

\begin{figure}[p!]
  \includegraphics[width=0.47\textwidth]{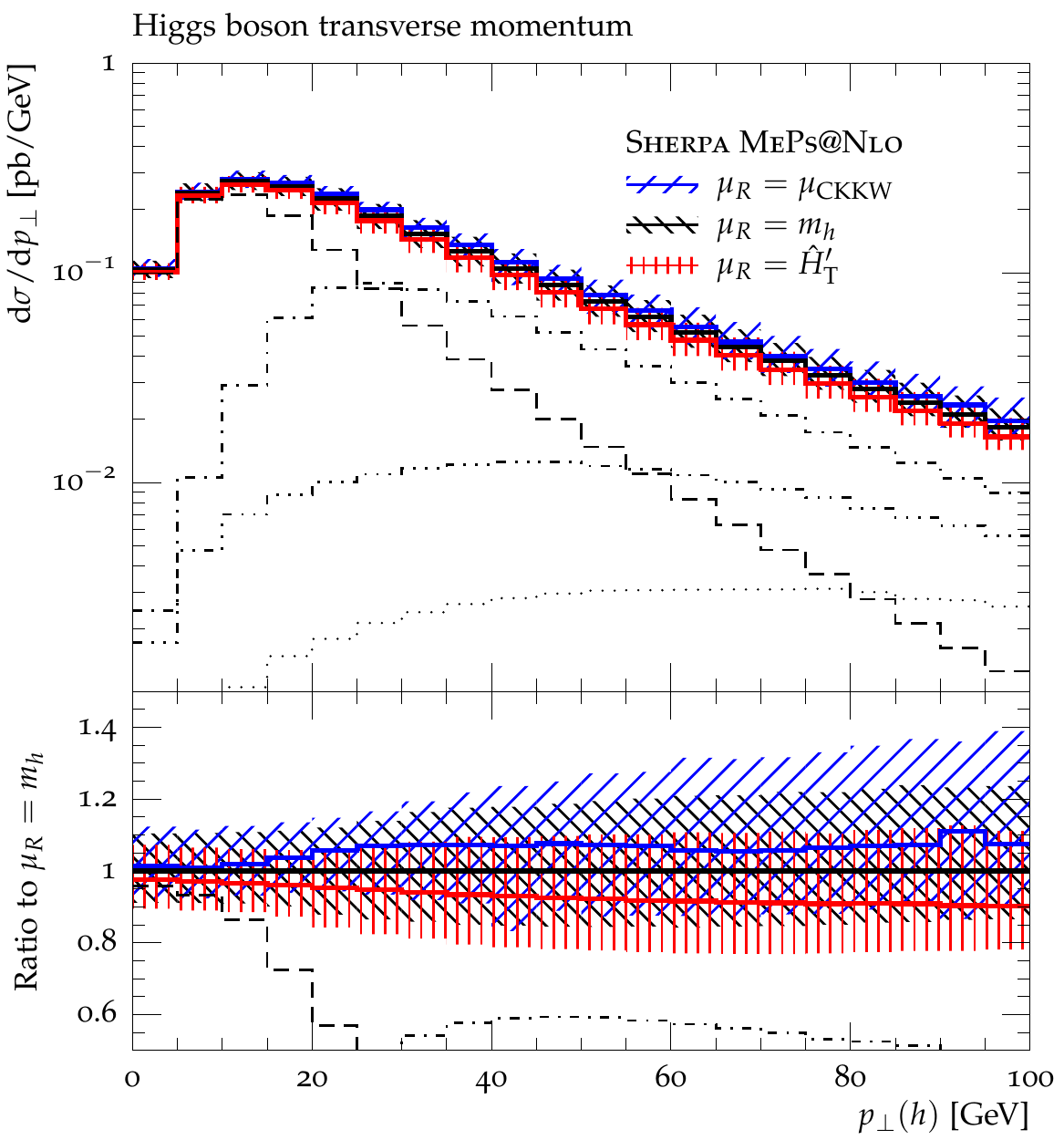}\hfill
  \includegraphics[width=0.47\textwidth]{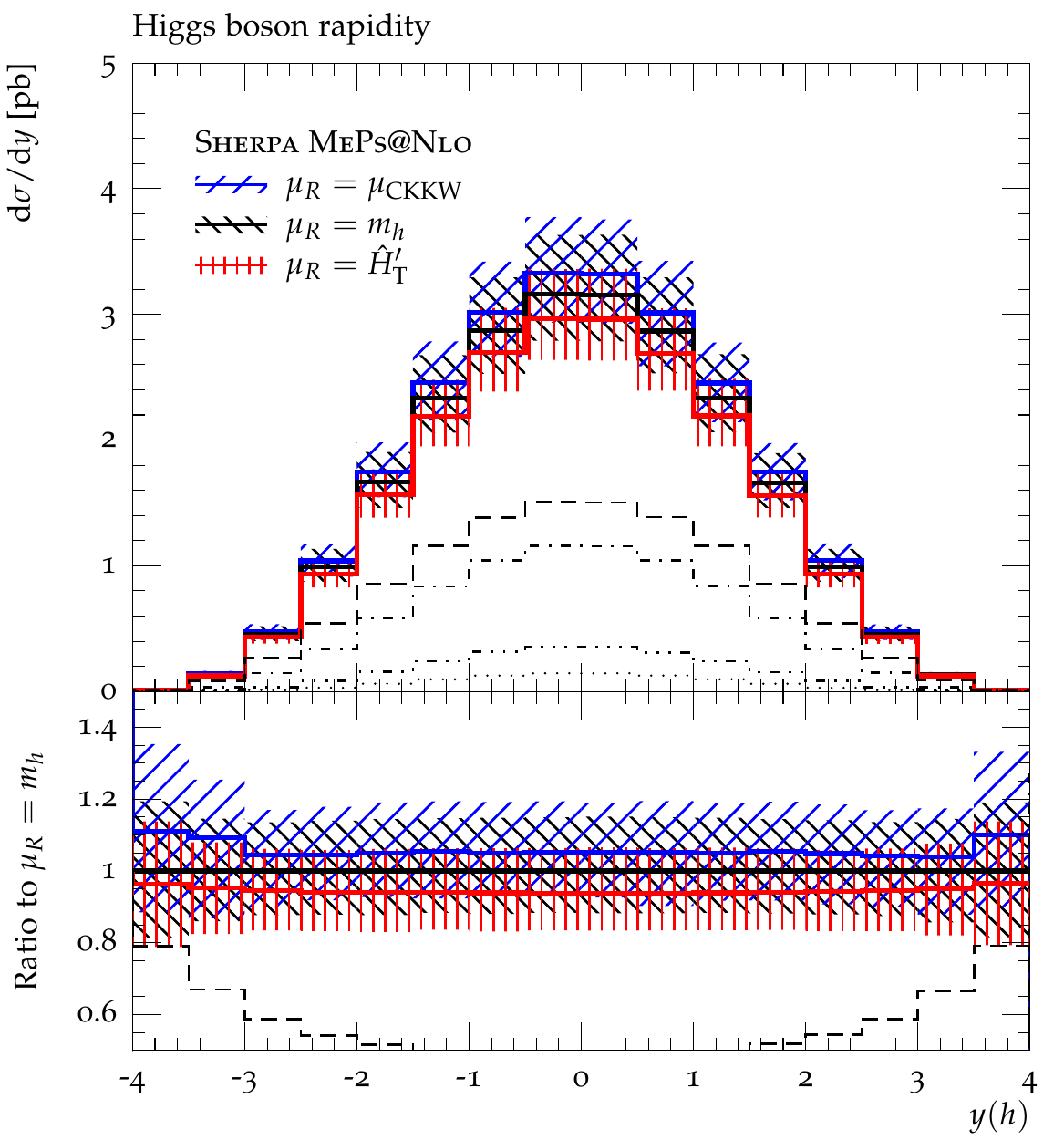}
  \caption{
           Transverse momentum (left) and rapidity (right) of the Higgs 
           boson 
           with three different scale choices. The uncertainty bands 
           include all sources of perturbative uncertainties as a quadratic sum.
           Dashed lines correspond to the contribution from $0$-jet hard scattering
           configurations, dashed-dotted lines to $1$-jet configurations,
           dotted-dashed lines to $2$-jet configurations and dotted lines to
           $3$-jet configurations.
  \label{fig:pth_yh_incl}}
\end{figure}

Next we analyze differential distributions. In Fig.~\ref{fig:pth_yh_incl} and following
the lower pane shows the ratio of all simulations with respect to the result from $\mu_R=m_h$.
The dashed line shows the contribution to the $\mu_R=m_h$ result from the exclusive NLO 
calculation for $pp\to h+0$ jets, and the dashed-dotted (dotted-dashed) line shows the contribution 
from the calculation for $pp\to h+1$ jet ($pp\to h+2$ jets). The dotted line shows the contribution 
from the LO result for $pp\to h+3$ jets. Uncertainty bands are computed as the quadratic sum of
the renormalization/factorization scale uncertainty in the perturbative calculation, the resummation
scale uncertainty and the multijet merging scale uncertainty. 
Intrinsic parton shower uncertainties will 
be discussed in detail in Sec.~\ref{sec:ps_unc}.

The transverse momentum distribution of the Higgs boson, and its rapidity spectrum 
are shown in Fig.~\ref{fig:pth_yh_incl}. Since both are rather inclusive observables,
we observe a good agreement of the predictions obtained with different scale choices.
The \CKKW scale produces a larger uncertainty band than the two others,
because at small $p_{T,h}$ it is driven by the transverse momentum of the first jet.
This leads to an increase in the inclusive cross section, while the rapidity distribution
of the Higgs boson is largely unaffected.

\begin{figure}[p]
  \includegraphics[width=0.47\textwidth]{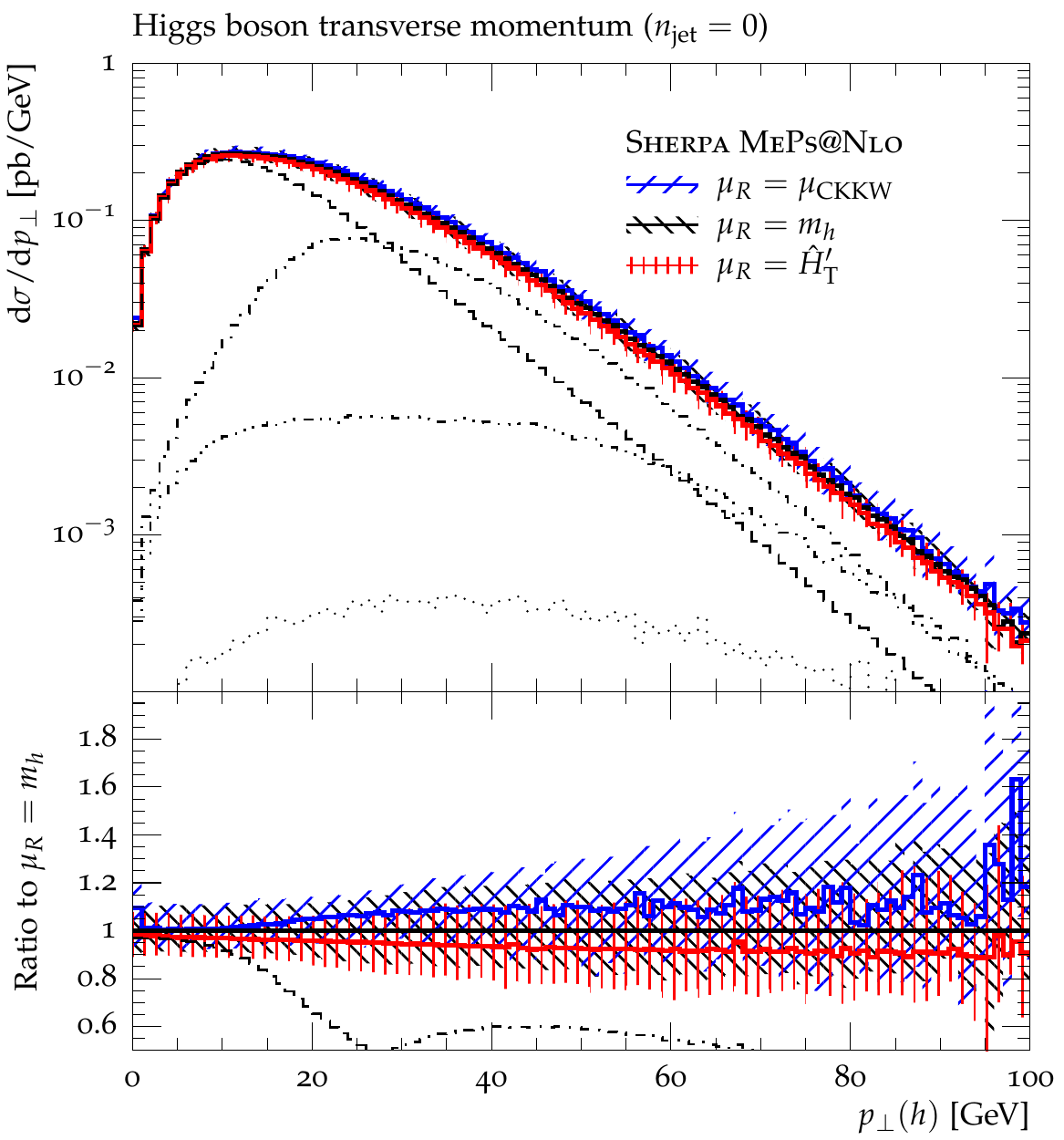}\hfill
  \includegraphics[width=0.47\textwidth]{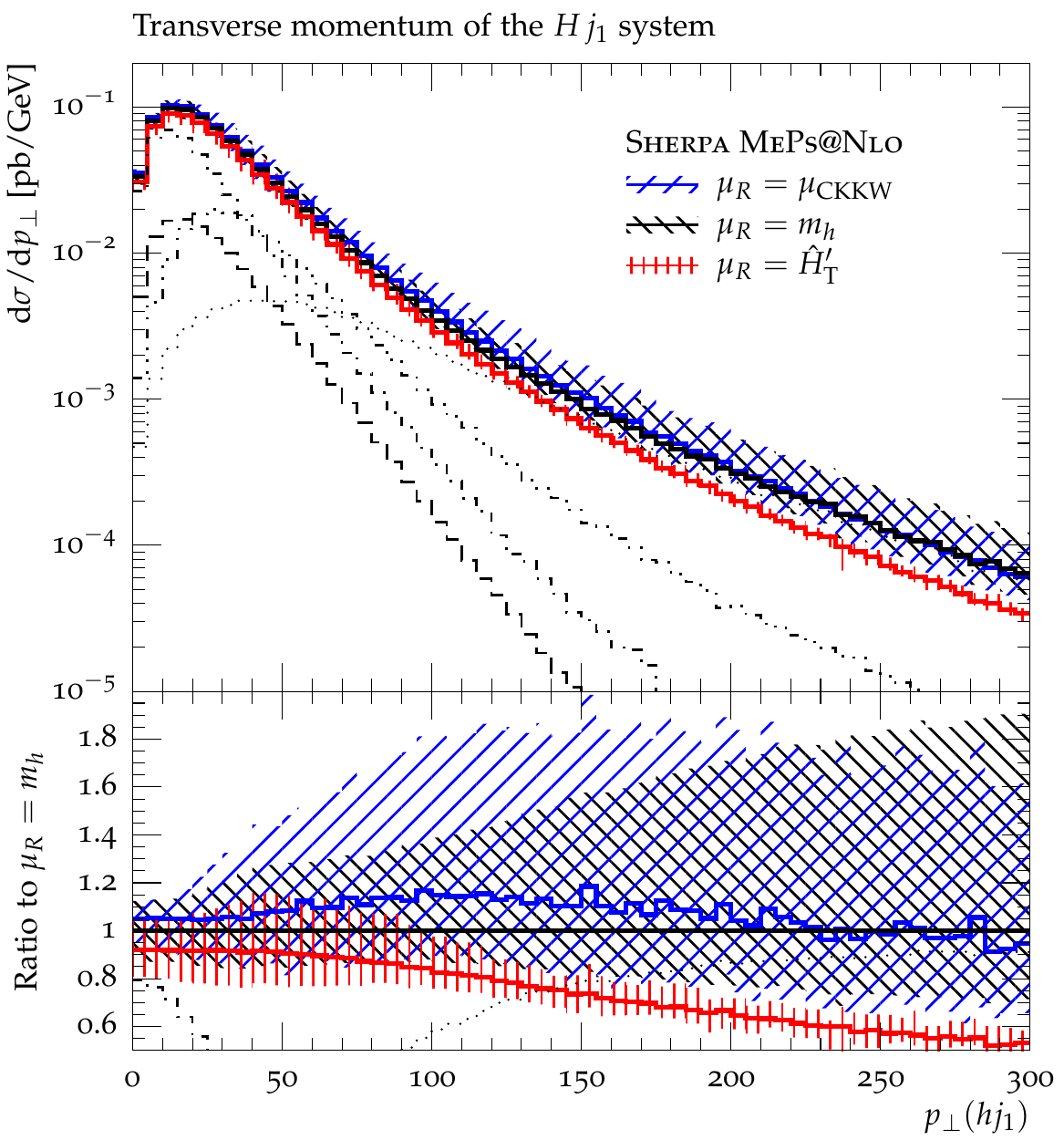}
  \caption{
           Transverse momentum of the Higgs boson in the absence (left) 
           and presence (right) of a jet with transverse momentum larger 
           than 30 GeV 
           with three different scale choices. For details see Fig.~\ref{fig:pth_yh_incl}.
  \label{fig:pth_jetveto_jetcut}}
\end{figure}

Figure~\ref{fig:pth_jetveto_jetcut} shows the Higgs transverse momentum distribution
in the absence of any jet with transverse momentum larger than 30~GeV, and the transverse
momentum distribution of the Higgs+jet system, $p_{T,hj}$, in the presence of a jet 
with $p_{T,j}>30$~GeV.
It is interesting to observe that the uncertainties in $p_{T,hj}$ are of similar size
at high transverse momentum for the \CKKW scale and for $\mu_R=m_h$. This is because
the \CKKW scale in this case is driven by the lowest scale in the parton-shower tree,
which is the scale of the core process, $m_h$.
Correspondingly, the uncertainties of the Higgs boson $p_T$ with a jet veto are 
of similar size for $\mu_R=m_h$ and for $\mu_R=\hat{H}_T^\prime$, even at large
transverse momenta. The difference in these scales is at most 
$\hat{H}_T^\prime-m_h\lesssim 120$~GeV, as we include up to only three additional jets 
at fixed order in the simulation.

\subsection{Dijet and VBF observables}
\label{sec:res-dijet-vbf}

\begin{figure}[p!]
  \includegraphics[width=0.47\textwidth]{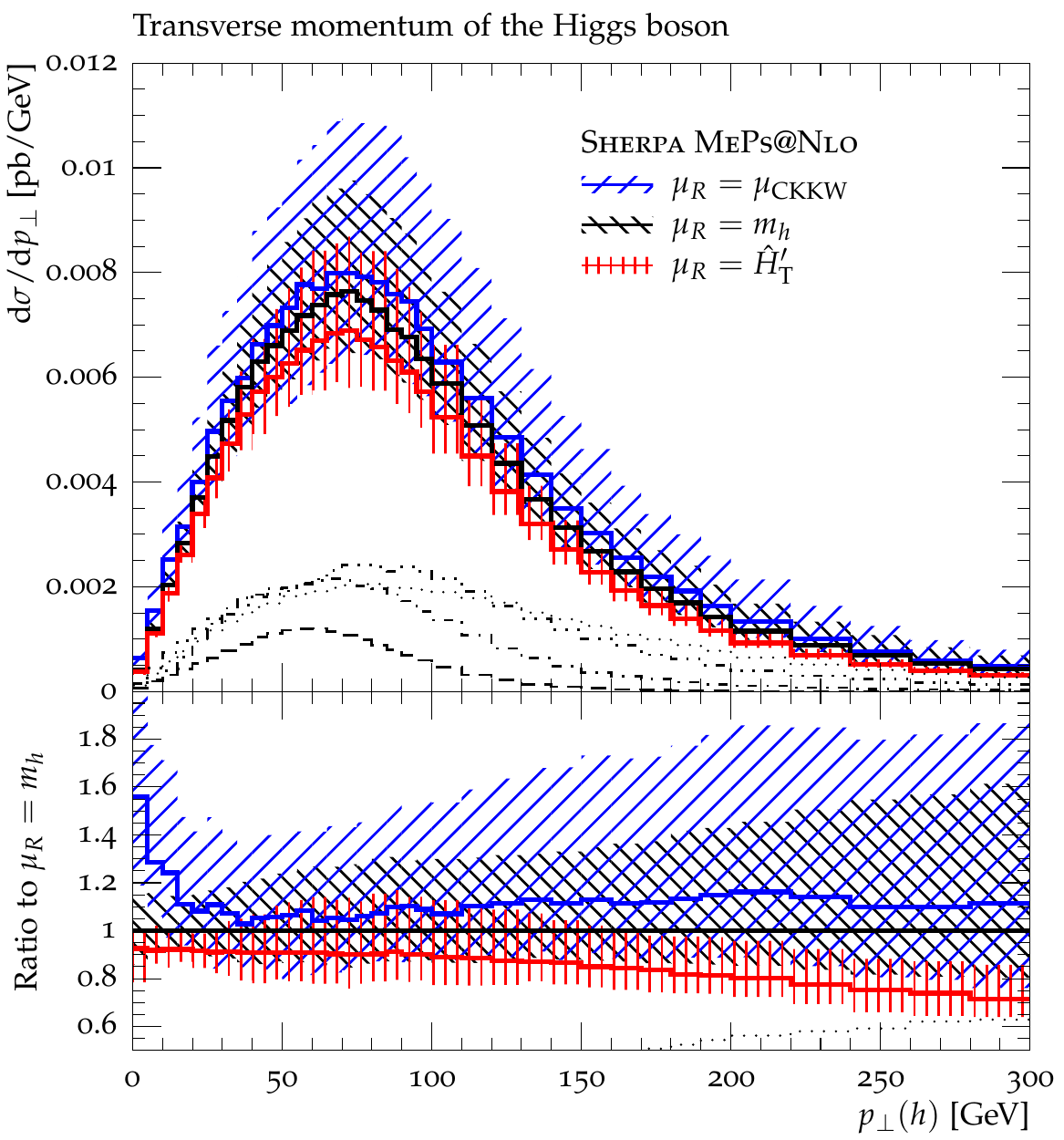}\hfill
  \includegraphics[width=0.47\textwidth]{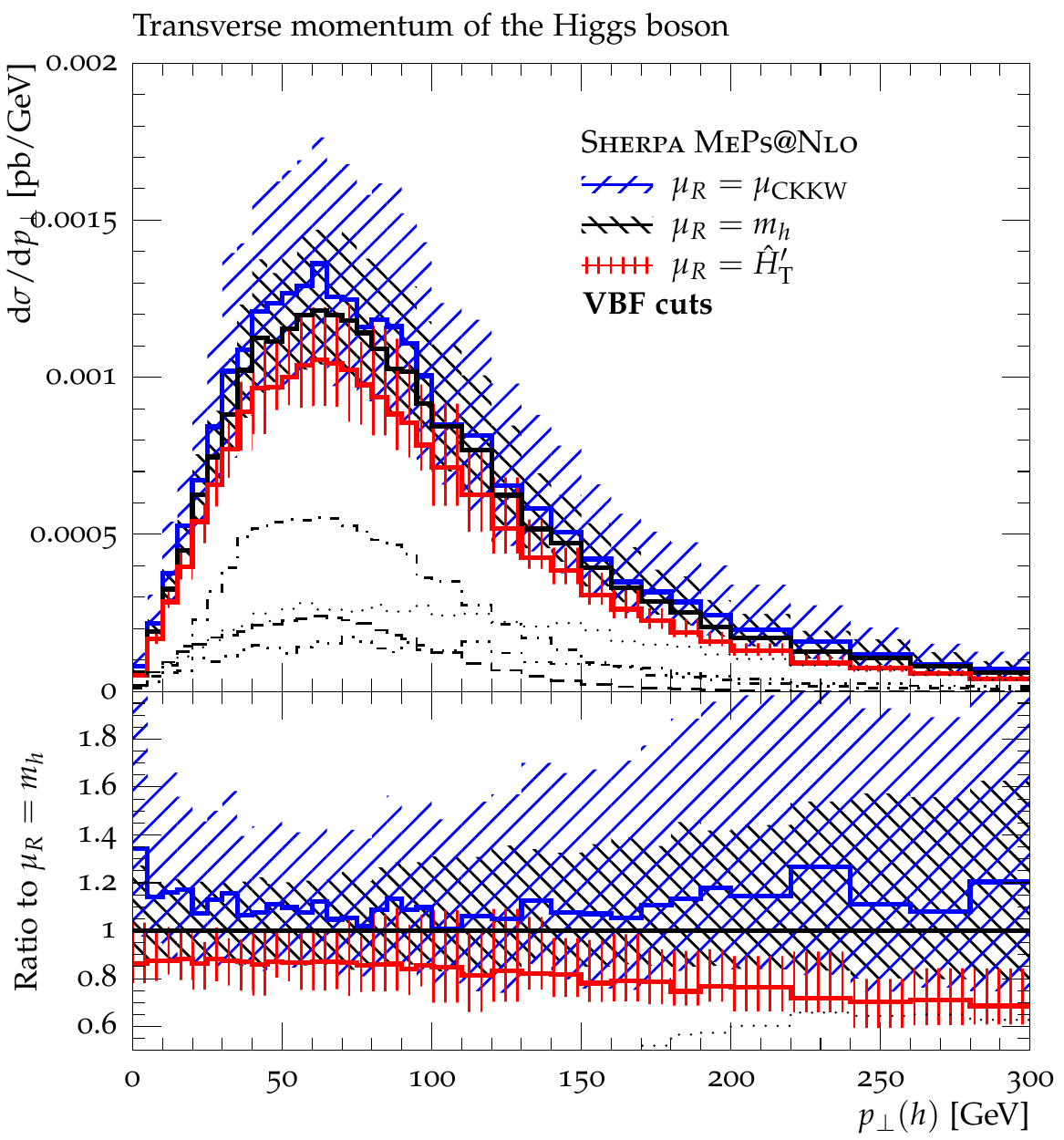}
  \caption{
           Transverse momentum of the Higgs boson in the dijet (left) and 
           VBF (right) selection 
           with three different scale choices. For details see Fig.~\ref{fig:pth_yh_incl}.
           \label{fig:pth_dijet_vbf}}
\end{figure}
\begin{figure}[p]
  \includegraphics[width=0.47\textwidth]{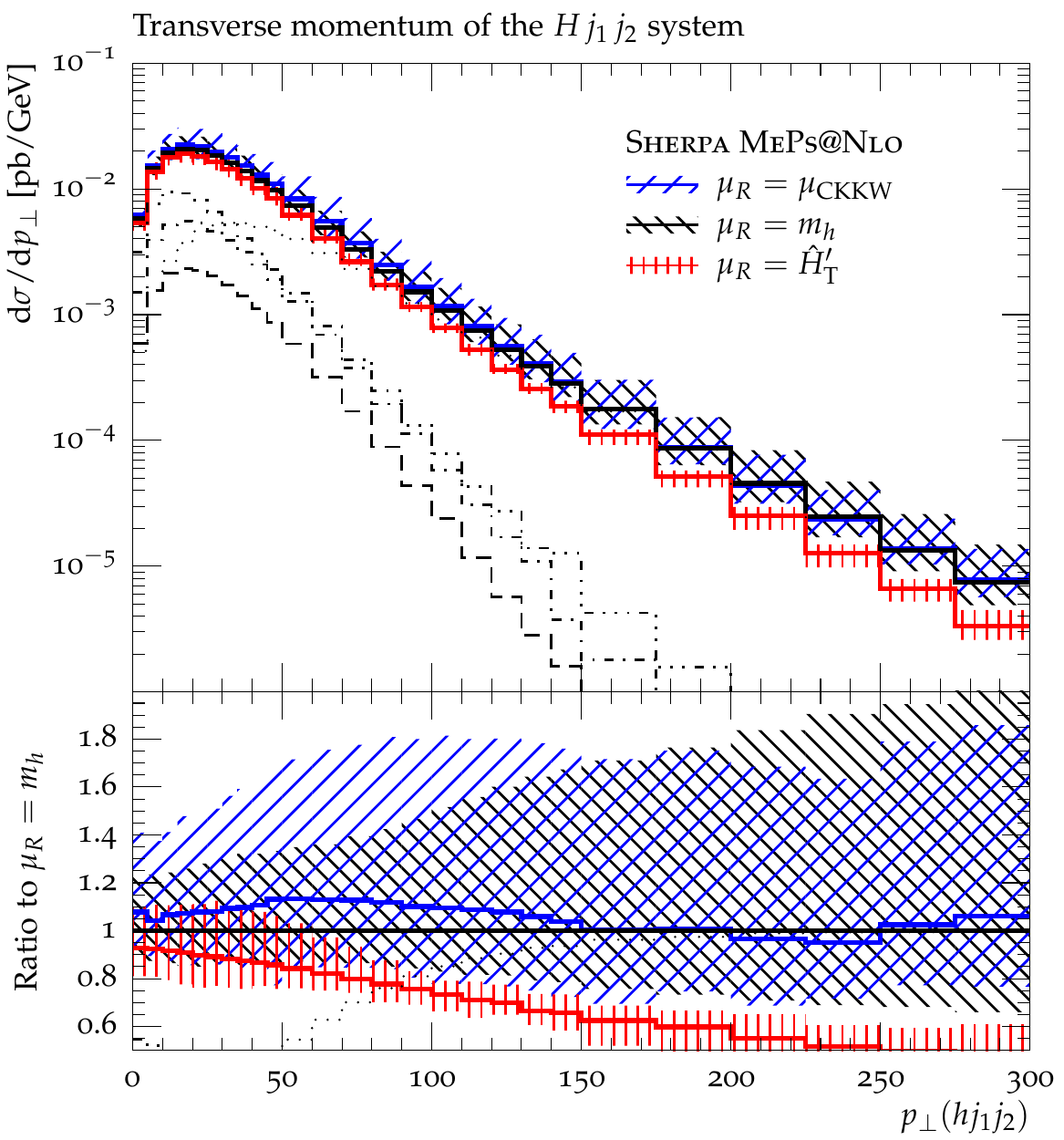}\hfill
  \includegraphics[width=0.47\textwidth]{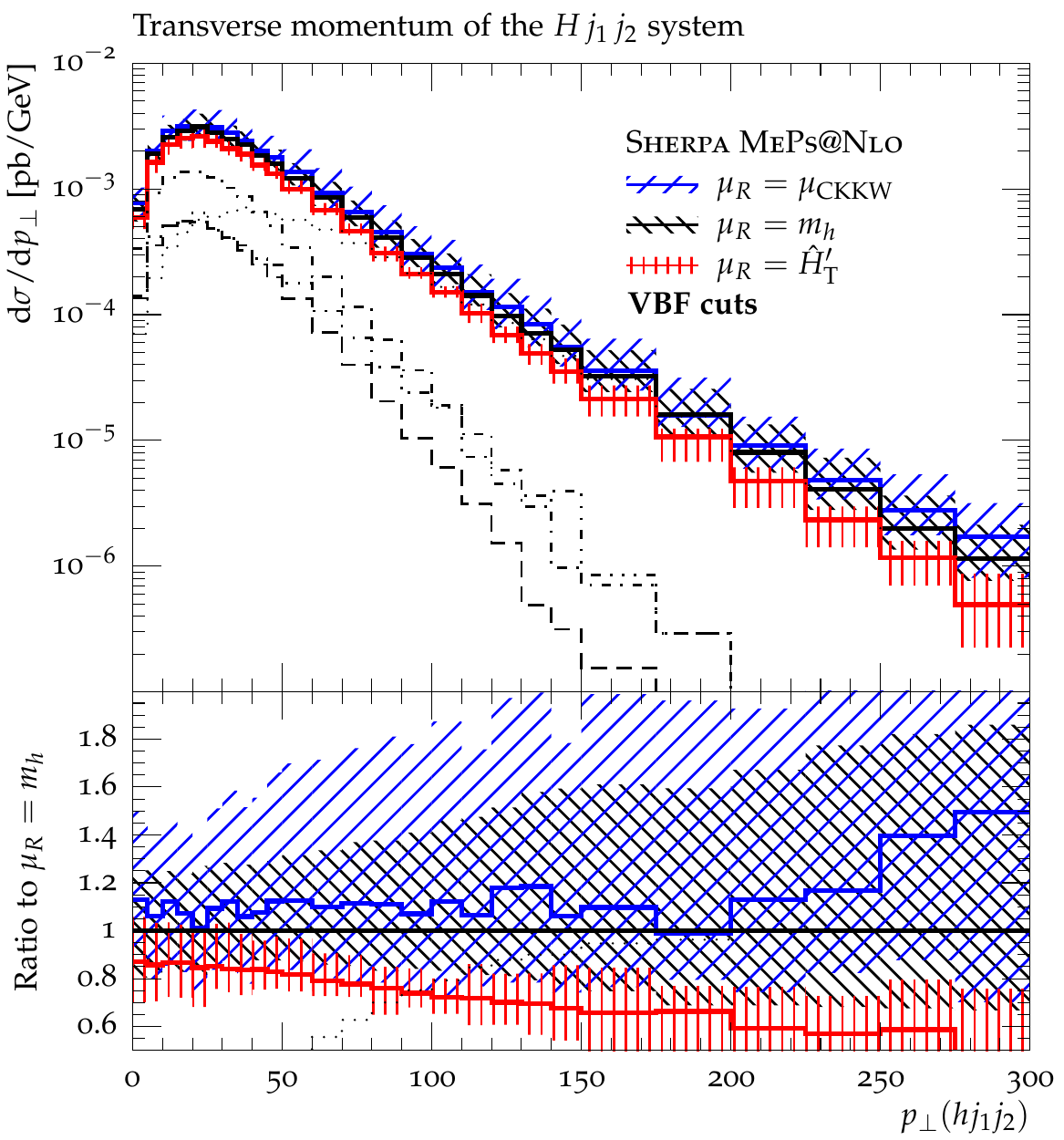}
  \caption{
           Transverse momentum of the Higgs boson plus two leading jets 
           system in the dijet (left) and VBF (right) selection 
           with three different scale choices. 
           For details see Fig.~\ref{fig:pth_yh_incl}.
           \label{fig:pthjj_dijet_vbf}}
\end{figure}

\begin{figure}[p!]
  \includegraphics[width=0.47\textwidth]{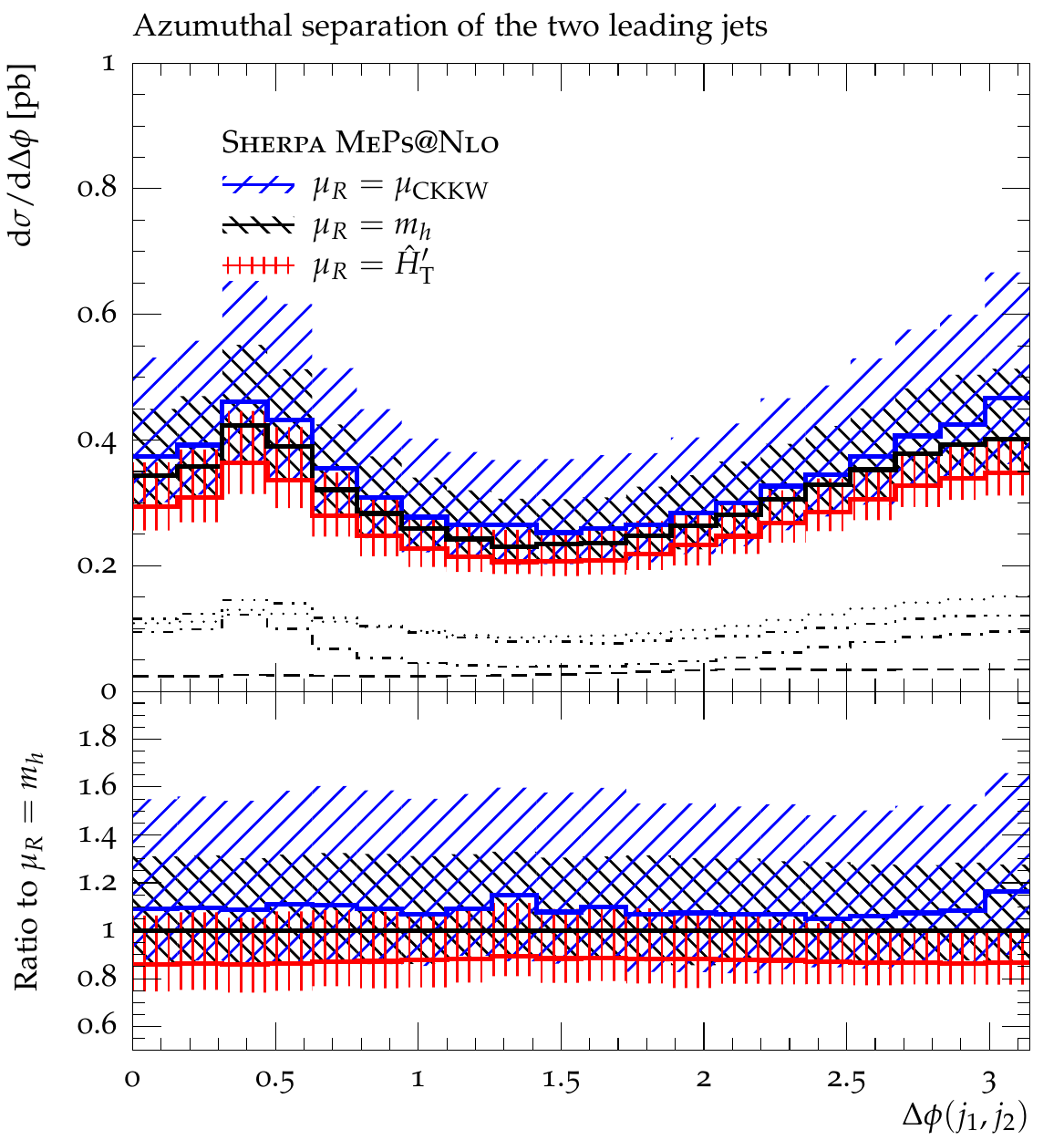}\hfill
  \includegraphics[width=0.47\textwidth]{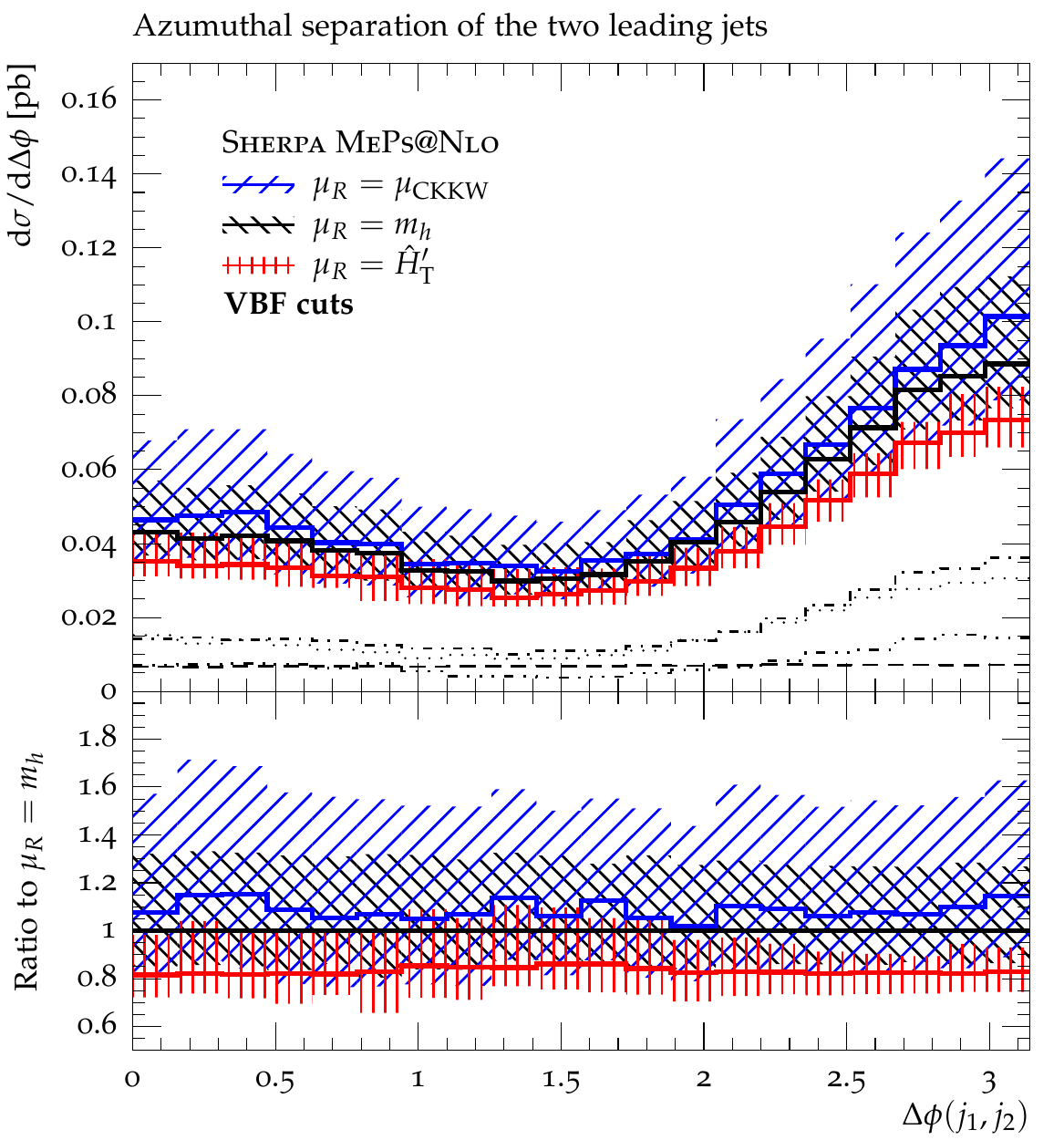}
  \caption{
           Azimuthal separation of the two leading jets in the dijet (left) 
           and VBF (right) selection 
           with three different scale choices.
           For details see Fig.~\ref{fig:pth_yh_incl}.
           \label{fig:dphijj_dijet_vbf}}
\end{figure}
\begin{figure}[p]
  \includegraphics[width=0.47\textwidth]{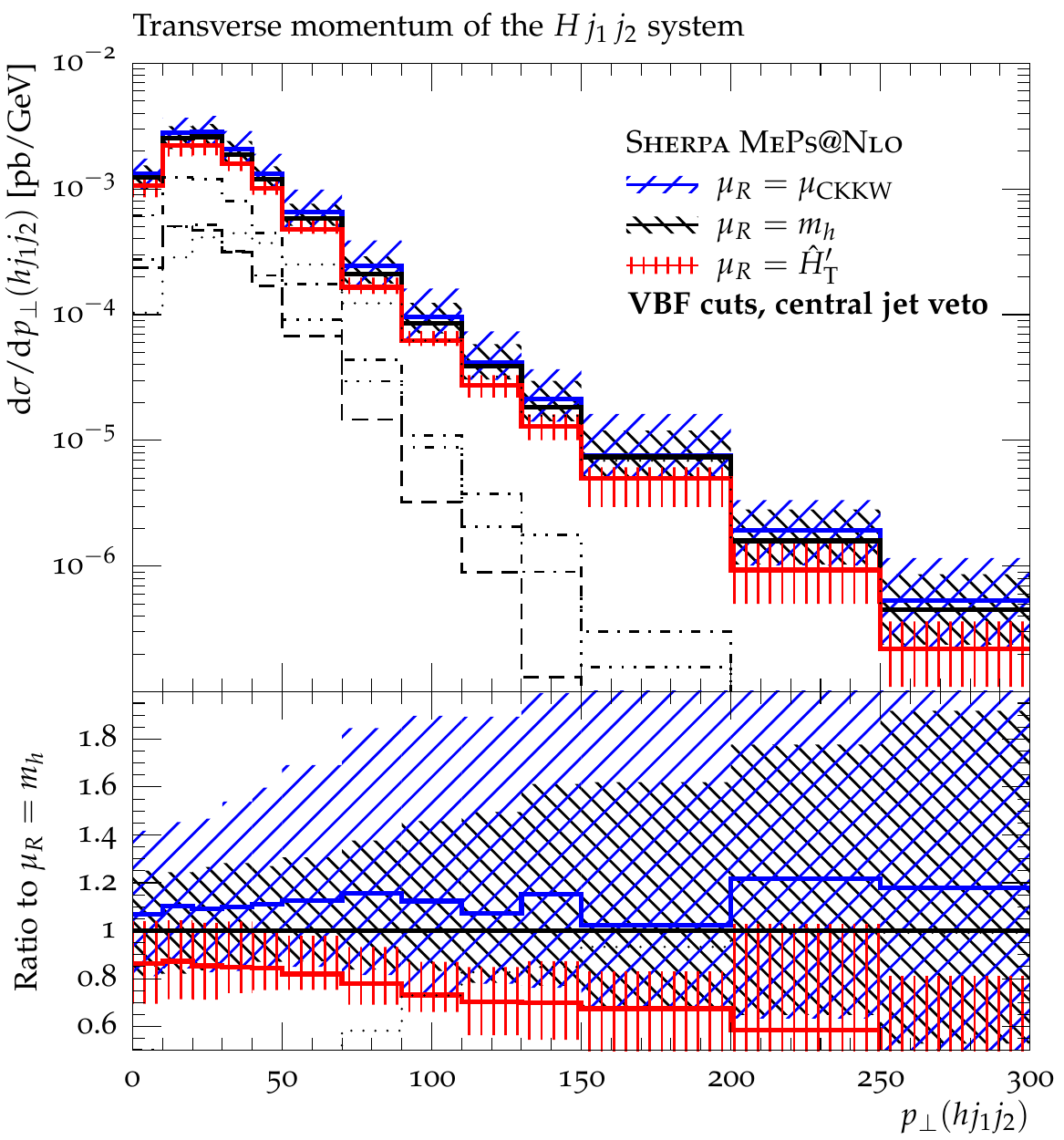}\hfill
  \includegraphics[width=0.47\textwidth]{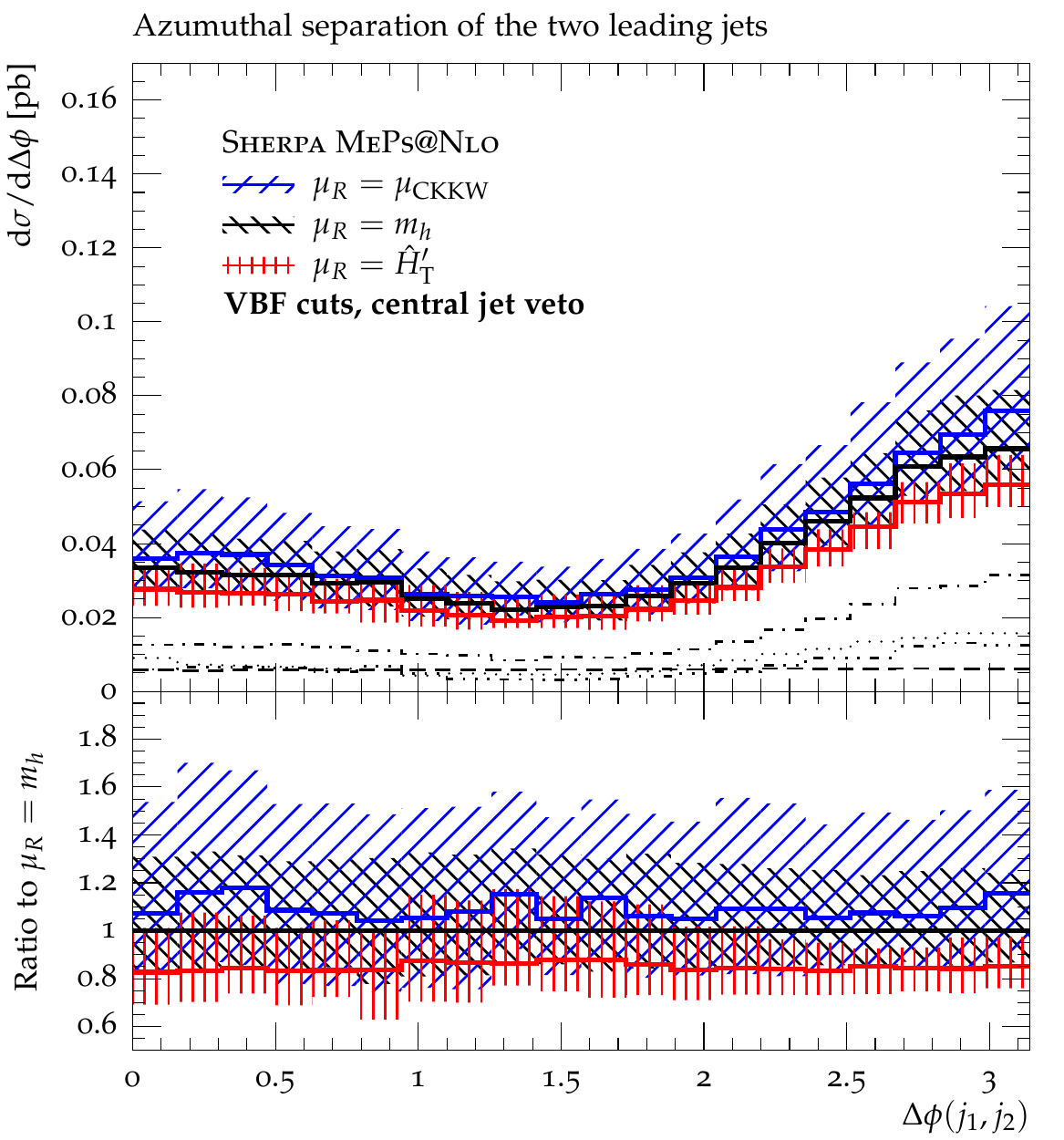}
  \caption{
           Transverse momentum of the Higgs boson plus two leading jets 
           system (left) and azimuthal separation of the two leading jets 
           (right) 
           with three different scale choices.
           For details see Fig.~\ref{fig:pth_yh_incl}.
           \label{fig:pthjj_dphijj_cntljveto}
          }
\end{figure}

Figure~\ref{fig:pth_dijet_vbf} shows the transverse momentum of the Higgs
boson in the dijet and the VBF selection. In the VBF selection the cross
section is considerably reduced, while the shape of the $p_T$ spectrum
stays largely the same. Note, however, that the VBF selection increases
the contribution from fixed-order events with exactly two hard jets,
indicated by the dashed-dotted line in the figure. This confirms that
also in the \MEPSatNLO merged sample the VBF selection acts as an effective
veto against a third jet, which is -- in gluon fusion processes -- 
predominantly produced in the central region.

Figure~\ref{fig:pthjj_dijet_vbf} shows the combined transverse momentum 
of the Higgs boson and the two leading jets. It is apparent that the 
uncertainties for the \CKKW scale and for $\mu_R=m_h$ 
are nearly identical in the high-$p_T$ region, an effect which was also
observed in Fig.~\ref{fig:pth_jetveto_jetcut}. The low-$p_T$ region
shows a large spread between the different predictions, as they are driven
in this case by their very different behavior with regard to adding an 
additional jet at small transverse momentum: In the \CKKW scheme, the 
overall scale will be close to the transverse momentum of this jet, 
and therefore rather small. This leads to an increase in the cross section.
For $\mu_R=\hat{H}_T^\prime$ the jet-$p_T$ will increase the scale further,
and the cross section will drop. The prediction for $\mu_R=m_h$ lies inbetween.

Fig.~\ref{fig:dphijj_dijet_vbf} displays
the azimuthal decorrelation between the Higgs boson and the dijet system,
and the azimuthal decorrelation between the two leading jets, respectively.
These observables do not exhibit a great sensitivity to the scale choice
in the fixed-order calculation. The only variations come from a change in
the total rate for Higgs plus dijet production, affecting normalization
of the result and size of its uncertainty band, but not its functional form. 
It is interesting to note the effect of the VBF cuts on $\Delta\phi(j_1,j_2)$. 
In the plain dijet selection two maxima are apparent, one at the jet radius 
stemming from the two leading jets being produced collinearly, recoiling 
against the Higgs boson, and the other at $\Delta\phi=\pi$ stressing the 
importance dijet-like topologies with a rather soft Higgs produced centrally. 
While the latter configurations are enhanced by the VBF cuts, the former 
are suppressed.

With the introduction of a veto on jet production inbetween the two leading 
jets the shape of the $\Delta\phi(j_1,j_2)$ distribution remains largely 
unaffected. Only its cross section is reduced, as can be seen in 
Fig.~\ref{fig:pthjj_dphijj_cntljveto}. The transverse momentum of the 
Higgs-plus-dijet system on the other hand is softened as its driving 
force, the production of a third jet, is constrained.

\subsection{Parton shower uncertainties}
\label{sec:ps_unc}
\begin{figure}[p!]
  \includegraphics[width=0.47\textwidth]{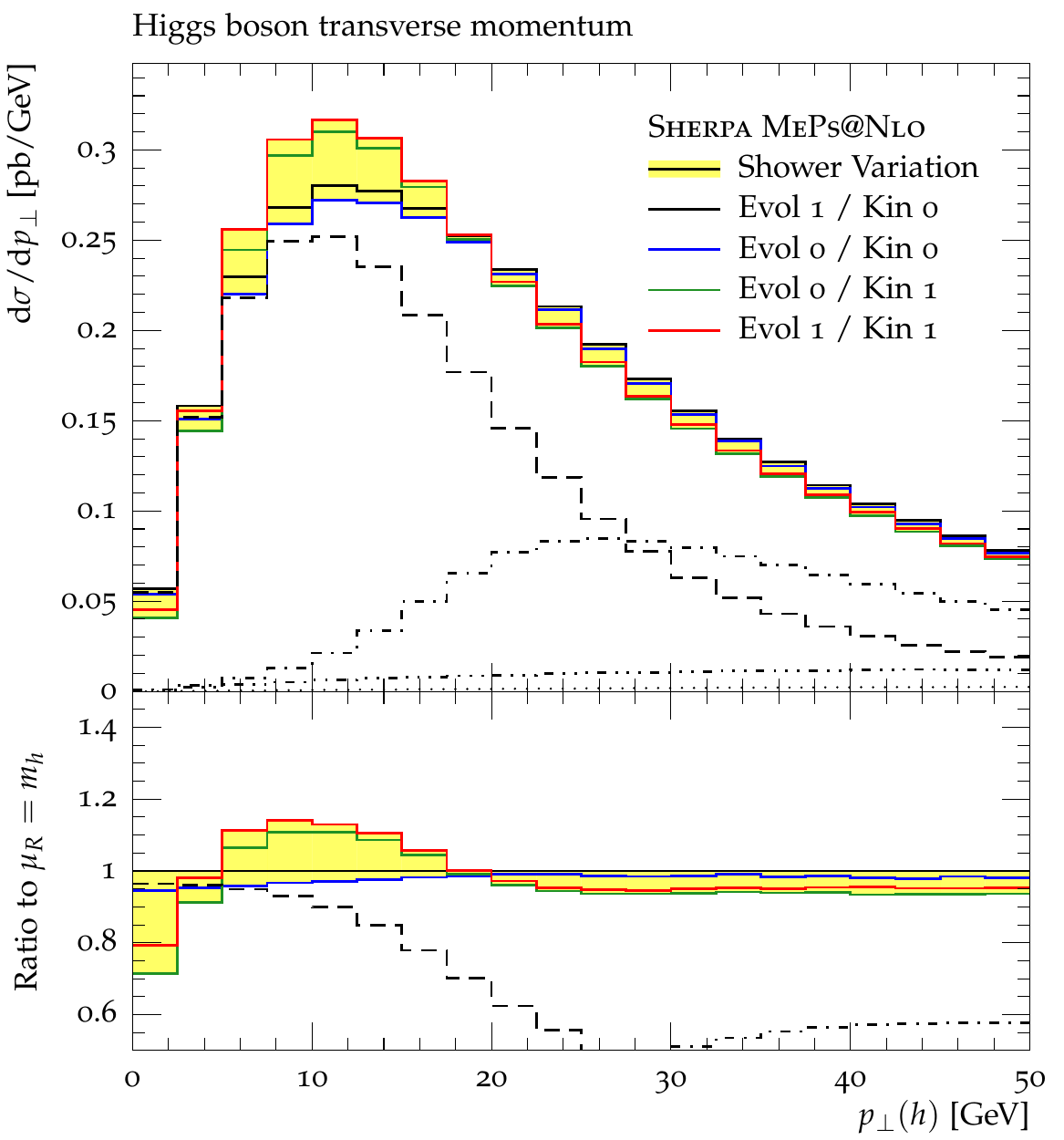}\hfill
  \includegraphics[width=0.47\textwidth]{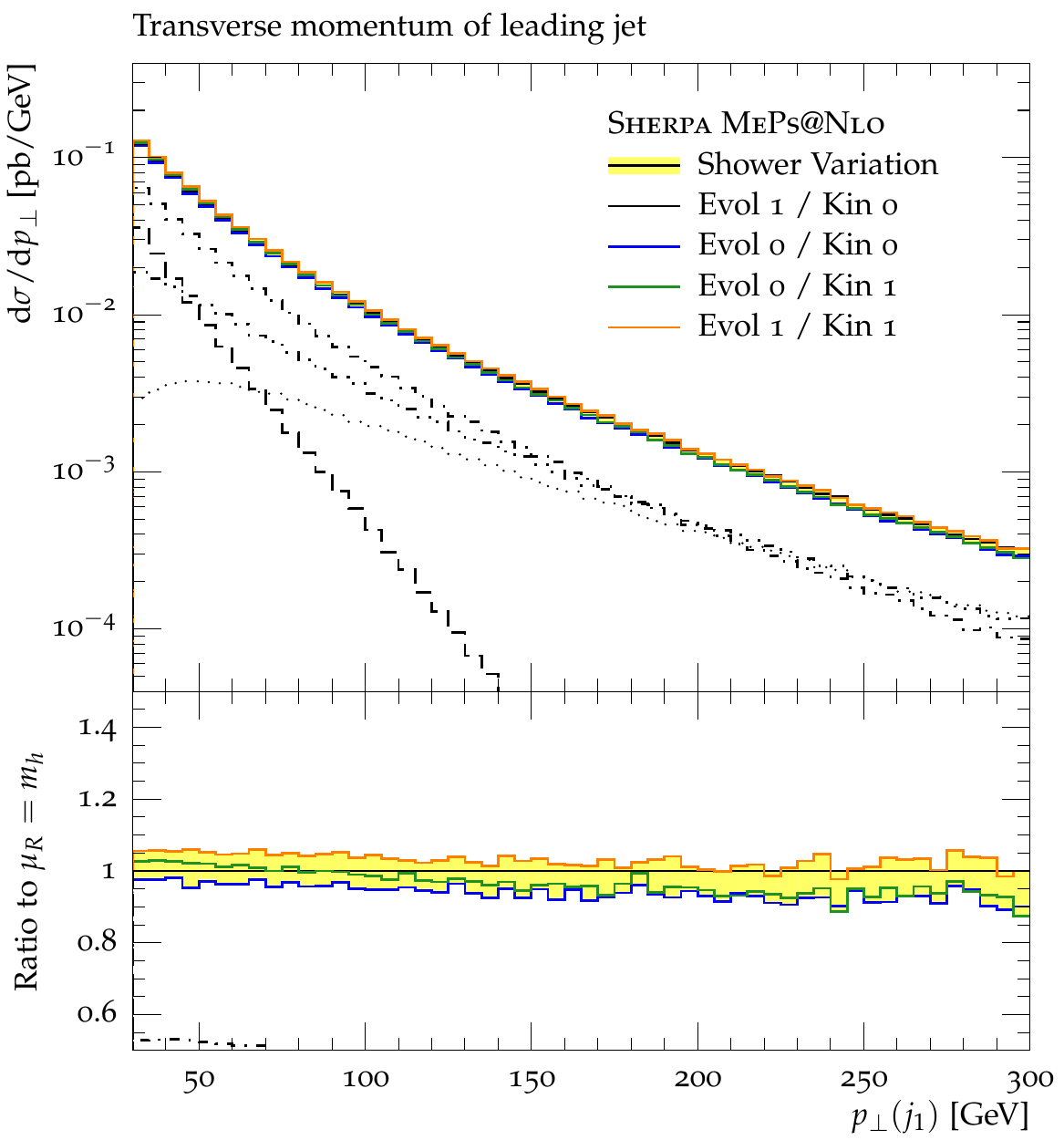}
  \caption{
           Transverse momentum of the Higgs boson (left) and of the first
           jet (right) for different evolution variables and recoil schemes.
           See Sec.~\ref{sec:methods} for details and definition of the schemes.
           \label{fig:psunc_pth_ptj}
          }
\end{figure}
\begin{figure}[p]
  \includegraphics[width=0.47\textwidth]{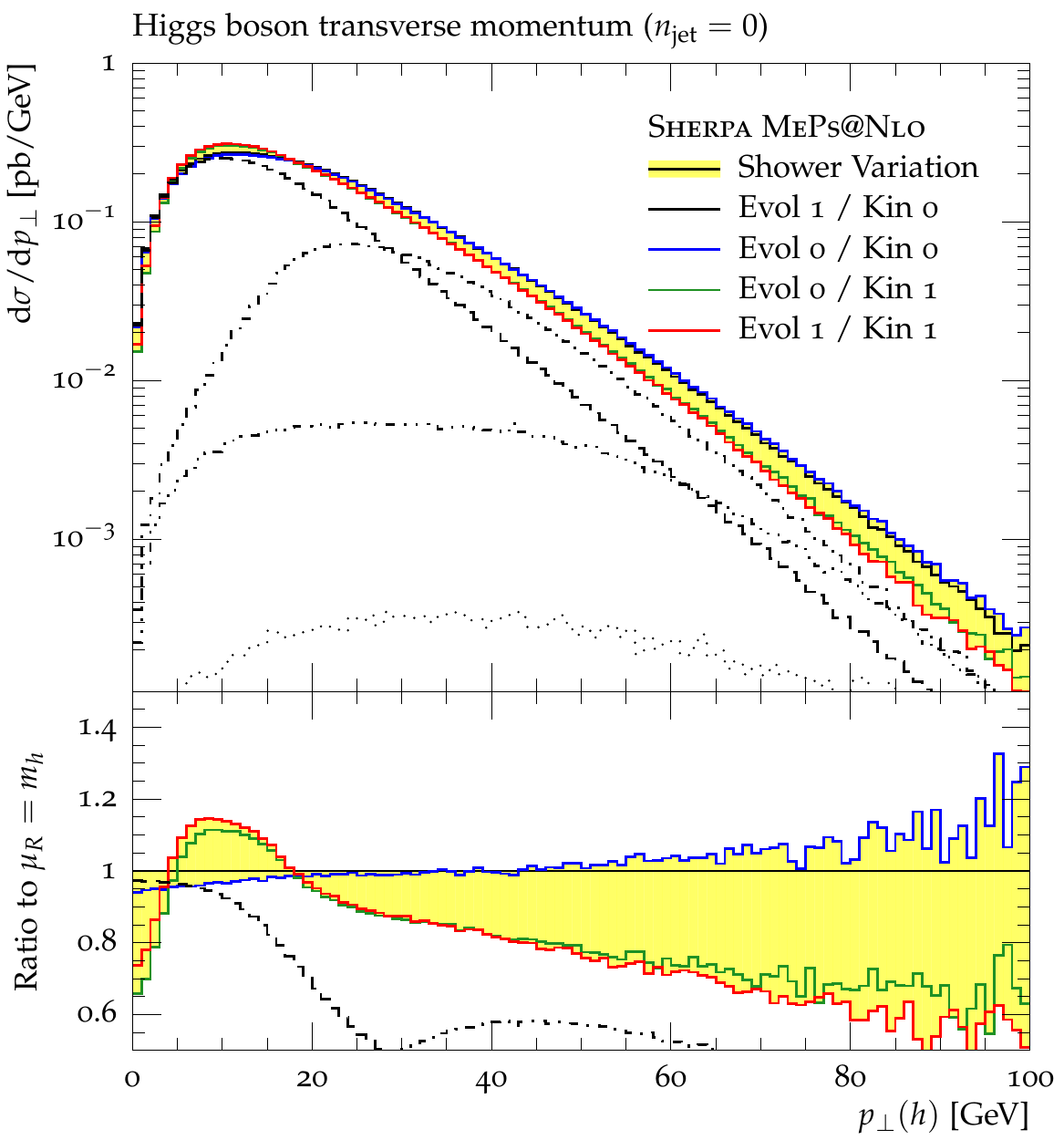}\hfill
  \includegraphics[width=0.47\textwidth]{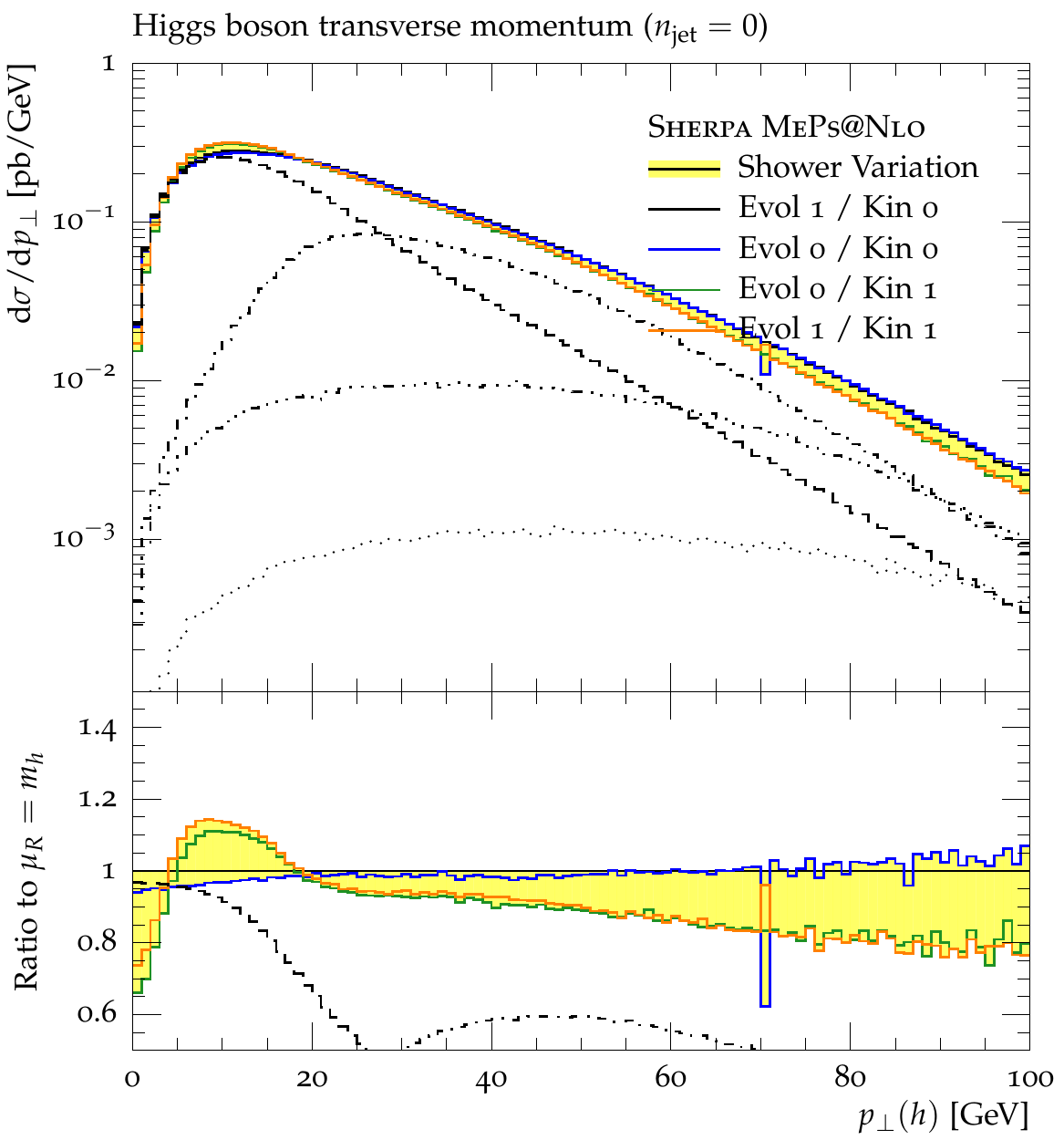}
  \caption{
           Transverse momentum of the Higgs boson in absence of jets with
           $p_T>30$~GeV (left) and $p_T>50$~GeV (right) for different 
           evolution variables and recoil schemes.
           See Sec.~\ref{sec:methods} for details and definitions of the schemes.
           \label{fig:psunc_pth_veto}
          }
\end{figure}

In this subsection the intrinsic uncertainties of the parton shower 
are scrutinized. We compare the two evolution schemes in Tab.~\ref{tab:evol_vars} 
and the two kinematics mappings described in Sec.~\ref{sec:methods}.

Figure~\ref{fig:psunc_pth_ptj} shows the transverse momentum of the Higgs boson
and the transverse momentum of the leading jet. We expect the dominant effects
of the kinematics mapping to be visible in the region which is most influenced 
by resummation. This is the low-$p_T$ region in the case of the Higgs transverse
momentum only. The transverse momentum of the hardest jet, by definition, should 
exhibit only a small sensitivity to resummation, which is nicely exemplified 
by a very small uncertainty band.

Next we turn to the transverse momentum of the Higgs boson in absence of any jet
of transverse momentum larger than 30~GeV (50~GeV), i.e.\ the transverse momentum
of the Higgs boson in presence of a jet veto.This observable must naturally be
extremely sensitive to the kinematics mapping in the high tail, because any
transverse momentum is generated by subsequent emissions of comparably low
transverse momentum. In other words, the large $p_T$ of the Higgs boson in this
case is built up by several mini-jets, predominantly produced by soft gluon emission
from initial-state partons. Figure~\ref{fig:psunc_pth_veto} shows that the 
uncertainty arising from the kinematics mapping is indeed dominant in the high-$p_T$
region, and that its size is considerably reduced when the cut on the jet-$p_T$
is increased. This is expected, because with a higher jet-$p_T$ cut the dominant
radiation effects again are modeled by emission of a few semi-hard jets, rather than
many soft gluons. This is also confirmed by the contributions from the various
individual $pp\to h+n$ jet results shown in Fig.~\ref{fig:psunc_pth_veto}.

\section{Conclusions}
\label{sec:conclusion}

We have presented an analysis of uncertainties in the merging of parton showers
with NLO QCD calculations for Higgs-boson plus multi-jets through gluon fusion.
We used the Monte-Carlo event generator \Sherpa for our study, in combination 
with virtual corrections obtained from \MCFM.

The uncertainties arising from the perturbative QCD calculation are sizable, 
due to the $\alpha_s^2$-dependence of the lowest multiplicity leading-order 
process.  This is reflected by the relatively large scale uncertainties, which
are driven by the variation of the renormalization scale.  The increased color 
charges in the initial state, compared to Drell-Yan processes, imply that 
resummation scale variations also have a much larger effect on the results.  
Additionally, the intrinsic parton-shower uncertainties, which we quantified 
through variations of the momentum mapping and the evolution parameter, have 
a large impact on observables involving a jet veto.

Consequently, the intrinsic uncertainties of the simulation are simultaneously 
driven by both the fixed-order part of the calculation and the resummation.
Effects which were observed in similar analyses of Drell-Yan lepton pair 
production, where deficiencies of the parton shower approximation could be 
largely removed by the fixed-order corrections are not as pronounced in 
Higgs-boson production.  Our analysis implies that more work is necessary to 
improve the resummation implemented by parton showers.  In particular, those uncertainties 
related to the definition of the momentum mapping pose an interesting problem
which probably can successfully be dealt with only by enhancing the 
accuracy of the shower approximation.  This should be seen in the context of striving
for higher accuracy in particle-level simulations through including
higher-order fixed order corrections.  In our
opinion, the potential improvement from next--\-to--\-next--\-to leading order 
corrections will be partially absorbed by the residual, potentially large 
uncertainties from parton showers, motivating a more inclusive approach.

\section*{Acknowledgments}

SH's work was supported by the US Department of Energy under contract 
DE--AC02--76SF00515 and in part by the US National Science Foundation, grant 
NSF--PHY--0705682, (The LHC Theory Initiative).  MS acknowledges supported 
by the Research Executive Agency (REA) of the European Union under the Grant 
Agreement number PITN-GA-2010-264564 (LHCPhenoNet) and PITN-GA-2012-315877 (MCnet).

MS would further like to acknowledge fruitful discussions on the subject 
within the \lq\lq Jets in Higgs physics\rq\rq{} working group for YR3 
\cite{Heinemeyer:2013tqa} and at Les Houches 2013.

This research was performed using resources provided by the Open Science Grid,
which is supported by the National Science Foundation and the US 
Department of Energy's Office of Science~\cite{Pordes:2008zz,*Bauerdick:2012xd}.

\bibliographystyle{bib/amsunsrt_modp}
\bibliography{bib/journal}
\end{document}